\documentclass[10pt, a4paper]{article}

\usepackage{jheppub}
\usepackage{axodraw4j}
\usepackage{pstricks}

\usepackage{color}
\usepackage{subfigure}

\allowdisplaybreaks[4]

\newcommand{\einhalb}{ \frac{1}{2} }

\newcommand{\cA}{\ensuremath{\mathcal{A}}}

\newcommand{\cD}{\ensuremath{\mathcal{D}}} % Pfadintegralmass

\newcommand{\cM}{\ensuremath{\mathcal{M}}}
\newcommand{\cN}{\ensuremath{\mathcal{N}}}
\newcommand{\cO}{\ensuremath{\mathcal{O}}} % generischer Operator

\newcommand{\mss}[1]{ {\mbox{\scriptsize #1}} }

\newcommand{\mycomment}[1]{ } 
\preprint{TTK-13-21}

\title{\boldmath The real radiation antenna functions for $S\rightarrow Q\bar{Q}gg$ at NNLO QCD}
\author[a]{Werner Bernreuther,}
\author[b]{Christian Bogner}
\author[a]{and Oliver Dekkers}
\affiliation[a]{Institut f\"ur Theoretische Physik,
RWTH Aachen University,\\
D-52056 Aachen, Germany}
\affiliation[b]{Institut f\"ur Physik,
Humboldt-Universit\"at zu Berlin, Unter den Linden 6,\\
D-10099 Berlin, Germany}
\emailAdd{breuther@physik.rwth-aachen.de}
\emailAdd{bogner@math.hu-berlin.de}
\emailAdd{dekkers@physik.rwth-aachen.de}
\abstract{
We analyze, in the 
 antenna subtraction framework, 
  the real radiation antenna functions for processes involving the 
production of a pair of heavy quarks and two gluons by an uncolored initial 
state at NNLO QCD. We provide explicit expressions for 
 these functions and discuss their 
infrared singular behaviour. Our main results are the corresponding integrated 
antenna functions which are computed analytically. They are expressed in terms 
of harmonic polylogarithms.
}
\begin{document}
\maketitle 
%
% version 18.9.2013  WB
% Version 14.9.2013,  WB 
%
%----------------------------------------------
%	Introduction
%----------------------------------------------
%
%
\section{Introduction} 
\label{section:intro}
In this paper we consider the  production of a heavy quark antiquark
 pair, $Q\bar{Q}$,  by an uncolored initial state $S$ at order
 $\alpha_s^2$, i.e., at
 next-to-next-to-leading order (NNLO) QCD:
\begin{equation} \label{intro:eq1}
 S \to Q \ {\bar Q} \ + X \, .
\end{equation}
Reactions of the type   \eqref{intro:eq1} include
 $e^+e^- \to \gamma^*, Z^* \to  Q  {\bar
Q} X$,  heavy quark-pair production by photon-photon collisions and
 by the decay of a colorless electrically neutral massive boson of
any spin. Our aim is to construct, within the antenna
 framework \cite{Kosower:1997zr,Kosower:2003bh,GRGG1},
  subtraction terms for regularizing
  and handling the infrared (IR) divergences that appear in the matrix
  elements which contribute to \eqref{intro:eq1} at NNLO QCD, such that the
  differential cross sections of \eqref{intro:eq1} can be computed to
  this order. Our paper is a second step in the
  computation of these NNLO subtraction terms: After having determined the
  subtraction term for the  $Q\bar{Q} q\bar{q}$ final state and its
  integral over the four-parton phase space in
   \cite{Bernreuther:2011jt}, we present here
  the analogous result for the  $Q\bar{Q} g g$ final state.
   
Before formulating the problem at hand, it seems appropriate to recall
 the state-of-the-art of the `subtraction technology' in perturbative QCD.
 For calculations at NLO QCD the presently most widely used approach is the
 dipole subtraction method  
\cite{Catani:1996vz,Catani:2000ef,Phaf:2001gc,Catani:2002hc}
 and  slight modifications  thereof \cite{Nagy:1998bb,Campbell:2004ch,Czakon:2009ss,Bevilacqua:2009zn,Frederix:2010cj,Goetz:2012uz}.
 A number of computer implementations of this method exist, including those of 
\cite{Gleisberg:2007md,Seymour:2008mu,Hasegawa:2009tx,Czakon:2009ss,
Frederix:2010cj}.
Other subtraction methods that were worked out at NLO QCD include 
 those of \cite{Frixione:1995ms,Nagy:1996bz,Frixione:1997np,Nagy:2003qn,Chung:2010fx,Chung:2012rq,Bevilacqua:2013iha}, and the  NLO subtraction within the antenna method  \cite{Kosower:1997zr,Kosower:2003bh,GRGG1,Daleo:2006xa,GehrmannDeRidder:2009fz,Abelof:2011jv}.

The infrared structure of 2-loop partonic amplitudes was analyzed 
in \cite{Campbell:1997hg,Catani:1999ss,Catani:2000pi,Bern:1998sc,Bern:1999ry,Kosower:2002su,Becher:2009qa,Becher:2009kw,Bierenbaum:2011gg}.
Techniques for handling the IR divergences of the individual contributions to partonic processes at NNLO QCD
 include the sector decomposition algorithm \cite{Binoth:2000ps,Binoth:2003ak,Binoth:2004jv,Anastasiou:2003gr,Carter:2010hi,Borowka:2012yc}, the antenna formalism \cite{GRGG1,Daleo:2009yj,Glover:2010im,Boughezal:2010mc,Gehrmann:2011wi,GehrmannDeRidder:2012ja,Currie:2013vh,Bernreuther:2011jt,Abelof:2012he,Abelof:2011ap, Abelof:2012rv}, 
 and the subtraction methods  \cite{Weinzierl:2003fx,Weinzierl:2003ra,Frixione:2004is,Somogyi:2005xz,Somogyi:2006da,Somogyi:2006db,Catani:2007vq,Somogyi:2008fc,Aglietti:2008fe,Bolzoni:2009ye,Bolzoni:2010bt}.
Applications to  reactions at NNLO QCD include  $pp \to H + X$ \cite{Anastasiou:2002yz,Anastasiou:2004xq},  
 $pp \to H + {\rm jet}$ \cite{Boughezal:2013uia},
 hadronic vector boson production \cite{Melnikov:2006di,Catani:2009sm,Catani:2011qz}, 
$pp \to H+W$   \cite{Ferrera:2011bk},  hadronic jet production \cite{Ridder:2013mf},
 $H\to b{\bar b} X$  \cite{Anastasiou:2011qx}, weak decays of heavy quarks \cite{Biswas:2009rb,Melnikov:2008qs,Gao:2012ja,Brucherseifer:2013iv},
$e^+ e^- \to 2 \ {\rm   jets}$   \cite{Anastasiou:2004qd,GehrmannDeRidder:2004tv}, 
and $e^+ e^- \to 3 \ {\rm jets}$
 \cite{GehrmannDeRidder:2007jk,GehrmannDeRidder:2008ug,Weinzierl:2008iv,Weinzierl:2009nz}.    
A general scheme, based on sector decomposition \cite{Binoth:2000ps,Binoth:2003ak} and phase-space partitioning 
  according to \cite{Frixione:1995ms},  that can be used
 for massless and massive partons, was presented in \cite{Czakon:2010td,Czakon:2011ve} and applied  in the computation
 of the total hadronic $t{\bar t}$ cross section to order $\alpha_s^4$ \cite{Baernreuther:2012ws,Czakon:2013goa}.
 For a related method and its application to $Z\to e^+e^-$, see  \cite{Boughezal:2011jf}.

At NNLO QCD the antenna method has been worked out completely so far only for  processes with massless partons in the final state.
 Our aim is to fill this gap for   reactions  of the type \eqref{intro:eq1}.
The theoretical description of the  reactions  \eqref{intro:eq1} to
 order  $\alpha_s^2$ requires i) the 2-loop amplitudes $S\to   Q 
{\bar Q}$.  They are known for $S =$ 
vector \cite{Bernreuther:2004ih,Gluza:2009yy}, axial vector 
\cite{Bernreuther:2004th,Bernreuther:2005rw},  scalar and pseudoscalar  
\cite{Bernreuther:2005gw}. ii)  The tree-level and one-loop amplitudes for $S \to  
 Q {\bar   Q} g$. The computation of the one-loop amplitudes is standard;
for instance, for $e^+e^-$ annihilation, i.e. $S=\gamma^*, Z^*$, 
they are given in \cite{Brandenburg:1997pu,Nason:1997nw,Rodrigo:1999qg}.
iii)  The tree-level
amplitudes $S\to  Q {\bar Q} Q {\bar Q},$  $Q {\bar Q} gg,$ and $Q {\bar Q} 
q {\bar q}$,  where $q$ denotes a massless quark. 

To be specific, let us discuss the  ingredients of a calculation 
of the differential cross section for $S$ decaying into two 
massive quark jets to order  $\alpha_s^2$. The computation of 
 $d\sigma_{\rm NLO}$ is standard.
 The contribution of order $\alpha_s^2$ to the 
two-jet cross section is given schematically by 
\begin{eqnarray}
d\sigma_{\mss{NNLO}} & = & 
\int_{\Phi_{4}} \! \left(d\sigma_{\mss{NNLO}}^{RR}-d\sigma_{\mss{NNLO}}^{S}
\right) 
+ \int_{\Phi_{3}} \! \left(d\sigma_{\mss{NNLO}}^{RV} - 
d\sigma_{\mss{NNLO}}^{T} \right) 
\nonumber \\
 & & {} + \int_{\Phi_{2}}d\sigma_{\mss{NNLO}}^{VV} 
+ \int_{\Phi_{4}} d\sigma_{\mss{NNLO}}^{S} 
+ \int_{\Phi_{3}} d\sigma_{\mss{NNLO}}^{T} \, .
\label{eq:cross_section_NNLO}
\end{eqnarray}
Here $d\sigma_{\rm NNLO}^{RR}$, $d\sigma_{\rm NNLO}^{RV}$, and
$d\sigma_{\rm NNLO}^{VV}$ denote the contributions from the 
 tree-level amplitudes\footnote{The IR-finite tree level
 amplitude $S\to  Q {\bar Q} Q {\bar Q}$ is of no concern here.}  $S\to Q {\bar Q} q {\bar q}$
and $S\to Q {\bar Q} g g$, the amplitude   $S\to Q {\bar Q} g$ to one-loop, 
and the amplitude $S\to Q {\bar Q}$ to two-loops,  respectively. 
 These individual contributions give rise to infrared (IR) divergences.
The remaining integrands of \eqref{eq:cross_section_NNLO}, $d\sigma_{\mss{NNLO}}^{S}$ and
$d\sigma_{\mss{NNLO}}^{T},$ are the double-real radiation subtraction
terms (for  $Q {\bar Q} gg$, $Q {\bar Q} q {\bar q}$)  and the real-virtual subtraction term, respectively. They have to be 
constructed such that the integrals over $d\sigma_{\mss{NNLO}}^{RR} - 
d\sigma_{\mss{NNLO}}^{S}$ and over $d\sigma_{\mss{NNLO}}^{RV} - 
d\sigma_{\mss{NNLO}}^{T}$ are finite and can be evaluated numerically. 
Furthermore, in order to make the cancellation of IR singularities explicit in 
eq.~\eqref{eq:cross_section_NNLO}, the integrals of these subtraction
 terms must be computed over the phase-space regions where IR singularities arise. 
  In the antenna subtraction
formalism, which we apply in the following, the subtraction terms
are constructed  from color-ordered matrix elements by
 exploiting  factorization properties of the 
 respective phase-space integrations.

Below, we determine  the color-ordered subtraction terms for
the $Q {\bar Q} g g$ final state 
from the matrix element of the process
\begin{equation} \label{intro:eq3}
 \gamma^*(q) \to Q(p_1) \ {\bar Q}(p_2) \ + g(p_3)\ g(p_4) \, .
\end{equation}
We then perform  the integrals of the subtraction terms
 analytically in  arbitrary space-time dimensions
over the full four-particle phase space. It is worth
 recalling that these (integrated) antenna
functions are not only of relevance 
 for the specific process at hand, but serve
also as building blocks for constructing subtraction terms for other processes
\eqref{intro:eq1} within the antenna formalism. 
  As mentioned above the (integrated)  subtraction term for
the $Q {\bar Q} q {\bar q}$ final state was already determined in   \cite{Bernreuther:2011jt}.
The remaining subtraction term required in \eqref{eq:cross_section_NNLO},
the  real-virtual subtraction term  
$d\sigma_{\mss{NNLO}}^{T}$, can be obtained from 
  the interference
  of the tree-level and 1-loop matrix element
  for  $\gamma^*\to  Q {\bar Q} g$.
 The integral of this term over the 3-particle phase space
 can also be done, in arbitrary space-time dimensions, in analytical 
   form and will be given in
  a future publication \cite{WBOD}.
 
The paper is organized as follows. In Section \ref{sec::antenna_fct} we determine the two color-ordered antenna functions for  (\ref{intro:eq3}) and discuss their singular limits. In Section \ref{sec::Integrated_antenna} we present the results of the integration of  these functions over the four-parton phase space in $D$ space-time dimensions. Section \ref{sec::conclusion} contains a summary and outlook.

% version 18.9.2013 WB
% version 17.9.2013 OD
\newpage
\section{The \texorpdfstring{\boldmath $Q\bar{Q}gg$}{QQgg} final state and 
antenna subtraction at NNLO}
\label{sec::antenna_fct}

\label{sec::2::1::intro}

As mentioned above, we will construct a subtraction term that coincides 
with the order $\alpha_s^2$ squared matrix element of
\begin{equation}
S ( q ) \to Q (p_1)\, \bar{Q} (p_2)\, g(p_3)\, g(p_4)
\end{equation}
in all single and double unresolved limits. The corresponding 
squared tree-level matrix element, summed over all colors and spins, but excluding the
symmetry factor $1/2$ for the two gluons in the final state, can be decomposed into color-ordered 
substructures as follows:
\begin{eqnarray}
%
% \lefteqn{
\big| M^0_{S \to Q \bar{Q} g g} \big|^2 
& = & \cN_0 \left( 4 \pi \alpha_s \right)^{2} \left( N_c^2 - 1 \right) 
% } \quad
%
\nonumber 
\\
& & {} \times
 \left[ N_c \left( \cM^0_4 ( 1_Q, 3_g, 4_g, 2_{\bar{Q}} ) 
+ \cM^0_4 ( 1_Q, 4_g, 3_g, 2_{\bar{Q}} ) \right)  
- \frac{1}{N_c}\, \tilde{\cM}^0_4 ( 1_Q, 4_g, 3_g, 2_{\bar{Q}} ) 
 \right] \,  , \qquad
\label{Antenna::MQQgg}
\end{eqnarray}
where $N_{c}$ denotes the number of colors. The normalization factor $\cN_0$ includes all non-QCD couplings. For the sake of brevity, we have dropped the dependence on the initial state momenta in the leading
 and subleading-color contributions $\cM^0_4$ and $\tilde{\cM}^0_4$. 
Adapting the notation of \cite{GRGG1}, we use symbolic labels $1_{Q},\,2_{\bar{Q}},\,3_{g},\,4_{g}$
  for the momenta of the quark, anti-quark and the gluons, respectively. 
In $\cM_{4}^{0}(1_Q, k_g, l_g, 2_{\bar{Q}} )$ the emission of the gluons $k$, $l$ is ordered, in the sense that there are 
color-connections between the  quark and gluon $k$, between the gluons $k$ and $l$, and between gluon $l$ and the antiquark. 
 In the subleading color term
  $\tilde{ \cM }^0_4$ both gluons are 
  photon-like, i.e., no non-abelian gluon vertices are involved.
Hence, when the two  gluons become collinear, this term does not become singular.

The corresponding contribution to the cross section for 2-jet production may be written as follows:
\begin{eqnarray}
d \sigma^{RR, Q \bar{Q} g g}_\mss{NNLO} & = & \einhalb\, \bar{\cN} \left( 4 \pi 
\alpha_s \right)^{2} \left( N_c^2 - 1 \right) d \Phi_4^{(D)} ( p_1,
p_2, p_3, p_4; q )\,  J^{(4)}_2 ( p_1, p_2, p_3, p_4 )
\nonumber \\
& & {} \times 
 \left[ N_c \left( \cM^0_4 ( 1_Q, 3_g, 4_g, 2_{\bar{Q}} ) 
+ \cM^0_4 ( 1_Q, 4_g, 3_g, 2_{\bar{Q}} ) \right)  
- \frac{1}{N_c}\, \tilde{\cM}^0_4 ( 1_Q, 4_g, 3_g, 2_{\bar{Q}} ) 
 \right] \, , \qquad
\label{Antenna::sigmaQQgg}
\end{eqnarray}
where the factor $1/2$ is due to Bose symmetry. 
The factor $\bar{\cN}$ contains $\cN_0$, the spin averaging factor for
 the initial state, and the  flux factor. 
 The phase-space measure of a $n$-parton final state in $D = 4 - 2 \epsilon$ dimensions is  given by
\begin{equation} \label{gm_3}
d \Phi^{(D)}_n ( p_1, \ldots , p_n ; q ) = \left( \mu^{ 4 - D } \right)^{n - 1} 
\prod^{n}_{i = 1} \frac{d^{D-1}p_{i}}{(2\pi)^{D-1}2p_{i}^{0}}  \left( 2 \pi 
\right)^{D}
\delta^{(D)}\! \left( q - \sum^n_{i = 1} p_i  \right), 
\end{equation}
where $\mu$ is a mass scale. The jet function $J^{(n)}_m$  in 
(\ref{Antenna::sigmaQQgg}) ensures that only configurations are taken into
account where $n$ outgoing partons form $m$ jets. 

For later reference, we define the charge and color stripped squared matrix elements 
 $\cM^0_2$ and $\cM^0_3$ associated with
the tree-level squared matrix elements of $S \to Q \bar{Q}$ and $S \to Q \bar{Q} g$, respectively:
\begin{eqnarray}
	\left| M_{S \to Q \bar{Q}} \right|^2 & = & \cN_0 N_c\, \cM^0_2 \! \left( 1_Q , 2_{\bar{Q}} \right) + \cO \!\left( \alpha_s \right) \,,
	\label{eq::M02_S2QQ}
	\\[1ex]
	\left| M_{S \to Q \bar{Q} g} \right|^2 & = & \cN_0 \left( 4 \pi \alpha_s \right) 
	\left( N_c^2 - 1 \right) \cM^0_3\! \left( 1_Q, 3_g, 2_{\bar{Q}} \right) + \cO \! 
	\left( \alpha_s^2 \right) \, .
	\label{eq::M03_S2QQg}
\end{eqnarray}
Summation over all spins  is understood.

%
% >>>>>>>>>>>>>>>>>>>>>>>>>>>>>>>>>>>>>>>>>>>>>>>>>>>>>>>>>>>>>>>>>>>>>>>>>>>>>>
%
\subsection{Antenna subtraction terms}
Let us now turn to the subtraction term $d\sigma_\mss{NNLO}^{S,\, Q\bar{Q}gg}$
 that must be constructed such that the phase space integration
over $d\sigma_\mss{NNLO}^{R,\, Q\bar{Q}gg}-d\sigma_\mss{NNLO}^{S,\, Q\bar{Q}gg}$
becomes finite in $D = 4$ space-time dimensions. We decompose this term into a sum of
two contributions, 
\begin{equation}
d\sigma_\mss{NNLO}^{S,\, Q\bar{Q}gg}=d\sigma_\mss{NNLO}^{S,\, a\; 
Q\bar{Q}gg}+d\sigma_\mss{NNLO}^{S,\, b\; Q\bar{Q}gg}, 
\end{equation}
where $d\sigma_\mss{NNLO}^{S,\, a\; Q\bar{Q}gg},\, d\sigma_\mss{NNLO}^{S,\, b\; 
Q\bar{Q}gg}$ cover the singularities due to single-unresolved and 
double-unresolved configurations, respectively. In analogy to the case of massless quarks \cite{GehrmannDeRidder:2007jk, Abelof:2011ap}, these terms are obtained as follows:
\begin{eqnarray}\label{sub::1}
\lefteqn{d \sigma^{S,a, Q \bar{Q} g g}_\mss{NNLO} = \einhalb\, \bar{\cN} \left( 4 \pi 
\alpha_s \right)^{2} \left( N_c^2 - 1 \right)\, 
d \Phi^{(D)}_4 ( p_1, p_2, p_3, p_4; q ) } \quad
\nonumber \\[1ex]
& & {} \times \bigg[ \, N_c \bigg(  d^0_3\! \left( 1_Q, 3_g, 4_g \right) 
\cM^{0}_3  \! \left( \widetilde{ (13) }_Q, \widetilde{ (43) }_g, 2_{\bar{Q}} \right) 
 J^{(3)}_2 \!  \left( \widetilde{ p_{13} }, \widetilde{ p_{43} }, p_2 \right) 
\nonumber \\[0.5ex]
& & \qquad \quad {} + d^0_3\! \left(2_{\bar{Q}}, 4_g, 3_g \right)
\cM^{0}_3  \! \left( 1_Q , \widetilde{ (34) }_g, \widetilde{(24)}_{\bar{Q}} \right) 
J^{(3)}_2 \! \left( p_1, \widetilde{ p_{34}}, \widetilde{ p_{24} } \right)
\nonumber \\[1ex]
& & \qquad \quad {} + d^0_3\! \left( 1_Q, 4_g, 3_g \right) 
\cM^{0}_3  \! \left( \widetilde{ (14) }_Q, \widetilde{ (34) }_g, 2_{\bar{Q}} \right) 
J^{(3)}_2 \!  \left( \widetilde{ p_{14} },\widetilde{ p_{34} }, p_2 \right) 
\nonumber \\[0.5ex]
& & \qquad \quad {} + d^0_3\! \left(2_{\bar{Q}}, 3_g, 4_g \right)
\cM^{0}_3  \! \left( 1_Q , \widetilde{ (43) }_g, \widetilde{(23)}_{\bar{Q}} \right) 
J^{(3)}_2 \! \left(  p_1, \widetilde{ p_{43}}, \widetilde{ p_{23} } \right) 
\bigg) 
\nonumber \\
& & \qquad {} - \frac{1}{N_c} \bigg( A^0_3 \! \left( 1_Q , 3_g , 2_{\bar{Q}} 
\right) 
\cM^{0}_3  \! \left( \widetilde{ (13) }_Q, 4_g, \widetilde{(23)}_{\bar{Q}} \right) 
J^{(3)}_2 \! \left( \widetilde{ p_{13}}, p_4, \widetilde{ p_{23} } \right)
\nonumber \\
& & \qquad \quad {} + A^0_3 \! \left( 1_Q , 4_g , 2_{\bar{Q}} \right) 
\cM^{0}_3  \! \left( \widetilde{ (14) }_Q, 3_g, \widetilde{(24)}_{\bar{Q}} \right) 
J^{(3)}_2 \! \left( \widetilde{ p_{14}}, p_3, \widetilde{ p_{24} } \right)
\bigg) \bigg],      
\\[4ex]
%
%\end{eqnarray}
%
% \begin{eqnarray}
%
\lefteqn{ d \sigma^{S,b, Q \bar{Q} g g}_\mss{NNLO} =
 \einhalb\, \bar{\cN} \left( 4 \pi 
\alpha_s \right)^{2} \left( N_c^2 - 1 \right)
\, d \Phi^{(D)}_4 ( p_1, p_2, p_3, p_4; q )}
 \quad
\nonumber \\[0.5ex]
& & {} \times \bigg[ \, N_c \bigg( A^0_4 \! \left ( p_1, p_3, p_4, p_2 \right) 
- 
d^0_3\! \left( 1_Q, 3_g, 4_g \right) 
A^{0}_3  \! \left( \widetilde{ (13) }_Q, \widetilde{ (43) }_g, 2_{\bar{Q}} 
\right) 
\nonumber \\[0.5ex]
& & \qquad {} - d^0_3\! \left(2_{\bar{Q}}, 4_g, 3_g \right) A^{0}_3 \! \left( 
1_Q , \widetilde{ (34) }_g, \widetilde{(24)}_{\bar{Q}} \right) \bigg)
\cM^{0}_2 \! \left( \widetilde{ ( 134 ) }_Q, \widetilde{ ( 234 ) }_{\bar{Q}} \right) 
J^{(2)}_2 \! \left( \widetilde{ p_{134} }, \widetilde{ p_{234} } \right) 
\nonumber \\[0.5ex]
& & \quad {} + N_c \bigg(  A^0_4 \! \left ( p_1, p_4, p_3, p_2 \right) - 
d^0_3\! 
\left( 1_Q, 4_g, 3_g \right) 
A^{0}_3  \! \left( \widetilde{ (14) }_Q, \widetilde{ (34) }_g, 2_{\bar{Q}} 
\right) 
\nonumber \\[0.5ex]
& & \qquad {} - d^0_3\! \left(2_{\bar{Q}}, 3_g, 4_g \right) A^{0}_3 \! \left( 
1_Q , \widetilde{ (43) }_g, \widetilde{ (23)}_{\bar{Q}} \right) \bigg) 
\cM^{0}_2 \! \left( \widetilde{ ( 143 ) }_Q, \widetilde{ ( 243 ) }_{\bar{Q}} \right) 
J^{(2)}_2 \! \left( \widetilde{ p_{143} }, \widetilde{ p_{243} } \right) 
\nonumber \\[0.5ex]
& &  \quad {} - \frac{1}{ N_c } \bigg( \tilde{ A }^0_4 \! \left( 1_Q, 3_g , 4_g 
, 2_{\bar{Q}} \right) - A^0_3 \! \left( 1_Q , 3_g , 2_{\bar{Q}} \right) A^0_3 
\! 
\left( \widetilde{ (13) }_Q , 4_g , \widetilde{ (23) }_Q \right)
\nonumber \\[0.5ex]
& & \quad {}  - A^0_3 \! \left( 1_Q , 4_g , 2_{\bar{Q}} \right) A^0_3 \! \left( 
\widetilde{ (14) }_Q , 3_g , \widetilde{ (24) }_Q \right) \bigg) 
\cM^{0}_2 \! \left( \widetilde{ ( 134 ) }_Q, \widetilde{ ( 234 ) }_{\bar{Q}} \right) 
J^{(2)}_2 \! \left( \widetilde{ p_{134} }, \widetilde{ p_{234} } \right) \bigg].
\label{sub::2}
\end{eqnarray}
The tree-level  massive quark-antiquark antenna
 function $A^0_3$ and the massive quark gluon antenna $d^0_3$ were 
derived in \cite{GehrmannDeRidder:2009fz}. The  four-parton  massive 
quark-antiquark antenna functions $A_{4}^{0}$ and $\tilde{A}_{4}^{0}$ govern the color-ordered and 
non-ordered (photon-like) emission of two gluons between a pair of massive radiator quarks, respectively. 
They constitute genuine NNLO objects that do not appear in the subtraction procedure at NLO. Their precise definition is given in eqs.~\eqref{eq::def::A40}, \eqref{eq::A40}, and \eqref{eq::A40t} below. The terms $\cM^0_2$, $\cM^0_3$ are defined in \eqref{eq::M03_S2QQg}, \eqref{eq::M02_S2QQ}, respectively. 
   
The subtraction terms \eqref{sub::1} and \eqref{sub::2} involve redefined on-shell momenta $\widetilde{p_{ij}}$ and $\widetilde{p_{ijk}}$, which are defined by Lorentz-invariant mappings $\{ p_i, p_j, p_k \} \to \{ \widetilde{p_{ij}}, \widetilde{p_{kj}} \}$ and $\{ p_i, p_j, p_k, p_l \} \to \{ \widetilde{p_{ijk}}, \widetilde{p_{ljk}} \}$. For massless final-state partons these mappings have been derived in ref.~\cite{GehrmannDeRidder:2007jk}. As discussed in ref.~\cite{Abelof:2011ap} the same mappings can be applied in the case of massive partons. 
%
%The momenta $\widetilde{p_{ij}}$ are linear combinations of $p_{i},\, p_{j}$. Likewise, the symbol  $\widetilde{(ij)}_{X}$ denotes a linear combination of momenta $i,\, j$ assigned to parton $X$.

An important feature of the antenna subtraction formalism is that
the infrared singular structure is governed by the integration over
individual antenna functions. The $(m+1)$-parton phase-space  measure can be factorized
 as follows:
\begin{equation}
d\Phi^{(D)}_{m+1}(p_{1},\ldots, p_{m+1};\, q) = d\Phi^{(D)}_{m}(p_{1},\ldots, \widetilde{p_{ij}},\, 
\widetilde{p_{kj}},\ldots, p_{m+1};\, q)\cdot d\Phi^{(D)}_{X_{ijk}}(p_{i},\, p_{j},\, p_{k};\, 
\widetilde{p_{ij}}+\widetilde{p_{kj}}).
\label{eq::phase-space-factorization}
\end{equation}
%
% where $P_{I},\, P_{K}$ are linear combinations of $p_{i},\, p_{j},\, p_{k}.$
Applying  \eqref{eq::phase-space-factorization} to the above subtraction terms and integrating
over the momenta not carrying a tilde in eq. \eqref{sub::1} one obtains
\begin{eqnarray}
\int_1 d \sigma^{S,a, Q \bar{Q} g g}_\mss{NNLO} & = & 
\bar{C}(\epsilon) \, 
\,  \bar{\cN} \left( 4 \pi \alpha_s \right)
\left( \frac{ \alpha_s }{ 2 \pi } \right) 
\left( N_c^2 - 1 \right)
d \Phi^{(D)}_3 ( p_1, p_2, p_3 ; q ) \, 
\cM^{0}_3  \! \left( 1_Q, 3_g, 2_{\bar{Q}} \right)
J^{(3)}_2 \!  \left( p_1 , p_{3} , p_2 \right) 
\nonumber \\
& & {} \times \left[ \, \frac{N_c}{2} \Big( \, 
  \cD^0_3\! \left( \epsilon; s_{13}, m_Q \right) 
+ \cD^0_3\! \left( \epsilon; s_{23}, m_Q \right) \Big) - \frac{1}{N_c} \cA^0_3 
\! \left( \epsilon, s_{12}, m_Q \right)  \right],
\label{sub::1_int}
\end{eqnarray}
where $\bar{C}(\epsilon) = 8 \pi^2 C( \epsilon ) = ( 4 \pi )^\epsilon 
e^{-\epsilon \gamma_E } $. Here and below we use the notation 
$s_{ij}=2p_{i}\cdot p_{j}$ and $s_{ijk}=s_{ij}+s_{ik}+s_{jk}$ for scalar 
products of momenta. The integrated massive antenna functions 
$\mathcal{A}_{3}^{0}$ and $\mathcal{D}_{3}^{0}$ are the integrals of $A_{3}^{0}$ 
and $D_{3}^{0}$ over appropriate phase-space regions, obtained from a 
factorization according to eq.~(\ref{eq::phase-space-factorization}). They 
were first computed in \cite{GehrmannDeRidder:2009fz} (see also eqs.~(5.15) and (5.17) of \cite{Abelof:2011jv}).

The subtraction term for the double unresolved configurations is split into two parts:
\begin{equation}
  d \sigma^{S,b, Q \bar{Q} g g}_\mss{NNLO} =
  d \sigma^{S,b,1, Q \bar{Q} g g}_\mss{NNLO} 
+ d \sigma^{S,b,2, Q \bar{Q} g g}_\mss{NNLO} \,,
\label{eq:sub12}
\end{equation}
where $d \sigma^{S,b,2, Q \bar{Q} g g}_\mss{NNLO}$ contains the terms that involve
the four-parton antenna functions $A^0_4$ and $\tilde{ A }^0_4$, while 
 $d \sigma^{S,b,1, Q \bar{Q} g g}_\mss{NNLO}$ contains all other terms. 
The subtracted contribution to the differential 2-jet cross section
\begin{equation}  \label{sbdifcrx}
\int_{\Phi^{(4)}_4} \left[ 
  d \sigma^{RR, Q \bar{Q} gg}_\mss{NNLO} 
- d \sigma^{S,a, Q \bar{Q} gg}_\mss{NNLO} 
- d \sigma^{S,b, Q \bar{Q} gg}_\mss{NNLO} 
\right]_{ \epsilon = 0 } 
\end{equation}
is  by construction IR finite  and can be integrated numerically over the four-parton phase space in $D=4$ dimensions. 

The  splitting \eqref{eq:sub12} is convenient because the 
 $\int d \sigma^{S,b,1, Q \bar{Q} g g}_\mss{NNLO}$ and $\int d 
\sigma^{S,b,2, Q \bar{Q} g g}_\mss{NNLO}$ are added  to 
 the three-parton and 
two-parton contribution to $d\sigma_{\mss{NNLO}}$, respectively (cf. \eqref {eq:cross_section_NNLO}).
   Hence they are  integrated over antenna 
phase spaces of different parton multiplicities. The respective integrals of \eqref{eq:sub12} read
\begin{eqnarray}
\int_2 d \sigma^{S,b,2, Q \bar{Q} g g}_\mss{NNLO} & = & 
\left( \bar{C} ( \epsilon ) \right)^2 
\bar{\cN} 
 \left( \frac{ \alpha_s }{ 2 \pi } \right)^2 
  \left( N_c^2 - 1 \right)
  \, 
d \Phi_2 ( p_1, p_2 ; q ) \,
\cM^{0}_{2} \! \left( 1_Q, 2_{\bar{Q}} \right) 
J^{(2)}_2 \!  \left( p_1 , p_2 \right)
\nonumber 
\\
& & { } \times \left[ 
N_c \, \cA^0_4 \! \left( \epsilon ; s_{12}, m_Q \right) 
- \frac{1}{2 N_c} \, 
\tilde{\cA}^0_4 \! \left( \epsilon ; s_{12}, m_Q \right) 
\right]\,,
\label{QQgg::sub::b::2::int}
\\[2ex]
\int_1 d \sigma^{S,b,1, Q \bar{Q} g g}_\mss{NNLO} & = & 
- \bar{C} ( \epsilon ) 
\,  \bar{\cN} \left( 4 \pi \alpha_s \right)
\left( \frac{ \alpha_s }{ 2 \pi } \right) 
\left( N_c^2 - 1 \right)
d \Phi_3 ( p_1, p_2, p_3 ; q ) \,
J^{(2)}_2 \!  \left( 
\widetilde{p_{13}}, 
\widetilde{p_{23}} 
\right)
\nonumber 
\\[1ex]
& & { } \times \cM^{0}_{2} \! \left( 
\widetilde{(13)}_Q, 
\widetilde{(23)}_{\bar{Q}} 
\right) 
 A^0_3 \!\left( 1_Q, 3_g, 2_{\bar{Q}} \right)  
\nonumber 
\\[1ex]
& & 
% \hspace{-15ex} 
{ } \times 
\left[ \, \frac{N_c}{2} \Big( \, 
  \cD^0_3\! \left( \epsilon; s_{13}, m_Q \right) 
+ \cD^0_3\! \left( \epsilon; s_{23}, m_Q \right) \Big) - \frac{1}{N_c} \cA^0_3 
\! \left( \epsilon; s_{12}, m_Q \right)  \right].
\label{QQgg::sub::b::1::int}
\end{eqnarray}
The integrated massive four-parton antenna functions $\cA^0_4$ and $\tilde{\cA}^0_4$, which have not been derived so far, will be given in sec.~\ref{sec::Integrated_antenna}.

%
%
% >>>>>>>>>>>>>>>>>>>>>>>>>>>>>>>>>>>>>>>>>>>>>>>>>>>>>>>>>>>>>>>>>>>>>>>>>>>>>>
%
\subsection{Antenna functions}
\label{sec::1::3::antennae}
We derive the
  antenna functions $A^0_4$ and $\tilde{A}^0_4$ 
  from the  color-ordered tree-level matrix element of the process
\begin{equation}
\gamma^\ast ( q ) \to Q (p_1)\, \bar{Q} (p_2)\, g(p_3)\, g(p_4) \, .
\end{equation}
The  respective squared matrix element, summed over spins and colors of the final state but excluding the Bose-symmetry factor $1/2$ reads 
% {\bf Kommentar: average of the gamma polarization, factor 1/2, included or not?}
%
\begin{eqnarray}
\big| {M}_{\gamma^\ast \to Q\bar{Q}gg}^{0} \big|^{2} 
& = & \left( 4 \pi \alpha \right) e_Q^2 \left( 4 \pi \alpha_s \right)^{2} \left( N_c^2 - 1 \right) \big| \cM^0_{\gamma^\ast \to Q\bar{Q}} \big|^2 
\nonumber
\\
& & {} \times \left[ N_{c} \left( 
  A_{4}^{0}( 1_{Q},\,3_{g},\,4_{g},\,2_{\bar{Q}} )
+ A_{4}^{0}( 1_{Q},\,4_{g},\,3_{g},\,2_{\bar{Q}} ) \right)
- \frac{1}{N_{c}}\tilde{A}_{4}^{0}( 1_{Q},\,4_{g},\,3_{g},\,2_{\bar{Q}} ) 
\right].
\nonumber
\\
\label{eq::def::A40}
\end{eqnarray}
The polarizations of $\gamma$ are summed, but not averaged. $e_Q$ denotes the electric charge of the massive quark in units of the positron charge $e = \sqrt{ 4 \pi \alpha }$ and
\begin{equation} \label{Antenna::MQQ}
\big| \cM^0_{\gamma^\ast \to Q\bar{Q}} \big|^2 = 4 \left[ \left( 1 - \epsilon \right) q^2 + 2 m_Q^2 \right] \, .
\end{equation}
The antenna functions $A^0_4$ and $\tilde{A}^0_4$ are given in appendix \ref{appendix::antennae}.
\subsection{Singular limits of the antennae}
\label{IRLimits}
Before integrating
 the  antenna functions $A^0_4$ and $\tilde{A}^0_4$, we study
  their behavior  in the single and double unresolved limits. Thereby 
 we verify  that the subtraction terms introduced in eqs.~\eqref{sub::1} and \eqref{sub::2} 
 lead  to a finite
 integral  \eqref{sbdifcrx} over the four-particle phase space. 
 The limiting behavior of the NLO antenna functions was already discussed
  extensively in the literature \cite{GehrmannDeRidder:2009fz,Abelof:2011jv}. Therefore we restrict ourselves to 
 the respective analysis of the above NNLO antenna functions.

We are not concerned with quasi-collinear limits \cite{Catani:2000ef} here, which will be relevant in numerical evaluations only if the squared mass of the quark $Q$ becomes much smaller than kinematic invariants.
\subsubsection{Single unresolved limits}
In the single unresolved limits, where one gluon becomes soft, the four-parton tree-level antennae $A^0_4$ and $\tilde{ A }^0_4$ factorise as 
%
% Soft limits 
%
\begin{eqnarray}
 A^0_4\! \left( 1_Q, i_g , j_g , 2_{\bar{Q}} \right) & \stackrel{ i_g \to 0 }{ 
\longrightarrow } & S \! \left( 1_Q, i_g, j_g \right) A^0_3 \! \left( 1_Q, j_g, 
2_{\bar{Q}} \right),
\\[1ex]
A^0_4\! \left( 1_Q, i_g , j_g , 2_{\bar{Q}} \right) & \stackrel{ j_g \to 0 }{ 
\longrightarrow } & S \! \left( 2_{\bar{Q}}, j_g, i_g \right) A^0_3 \! \left( 1_Q, i_g, 
2_{\bar{Q}} \right) , 
\\[1ex]
\tilde{A}^0_4\! \left( 1_Q, i_g , j_g , 2_{\bar{Q}} \right) & \stackrel{ i_g \to 0 }{ 
\longrightarrow } & S \! \left( 1_Q, i_g, 2_{\bar{Q}} \right) A^0_3 \! \left( 1_Q, j_g, 2_{\bar{Q}} \right) , 
\\[1ex]
\tilde{A}^0_4\! \left( 1_Q, i_g , j_g , 2_{\bar{Q}} \right) & \stackrel{ j_g \to 0 }{ 
\longrightarrow } & S \! \left( 1_Q, j_g, 2_{\bar{Q}} \right) A^0_3 \! \left( 1_Q, i_g, 2_{\bar{Q}} \right) , 
\end{eqnarray}
where 
\begin{equation} \label{SingleEikonal}
 S( i, j, k ) = \frac{ 2 s_{ik} }{ s_{ij} s_{jk} } - \frac{ 2 m_i^2 }{ s_{ij}^2 
} - \frac{ 2 m_k^2 }{ s_{jk}^2 }
\end{equation}
denotes the generalized single-soft eikonal function. 

When the two color-connected gluons become collinear, the antenna function $A^0_4$ behaves as follows:
%
% Soft limits 
%
\begin{equation}
A^0_4\! \left( 1_Q, i_g , j_g , 2_{\bar{Q}} \right) \: \stackrel{ i_g \parallel j_g }{ \longrightarrow } 
\frac{1}{s_{34}}\, P_{gg \to g} ( z )\,  A^0_3 \! \left( 1_Q, (ij)_g, 2_{\bar{Q}} \right) + \mbox{angular},
\label{A04coll}
\end{equation}
where $z$ is the momentum fraction carried by one of the collinear gluons and 
\begin{equation}
P_{gg \to g} ( z ) = 2 \left[
\frac{z}{1 - z} + \frac{ 1 - z }{z} + z ( 1 - z ) 
\right]
\end{equation}
is the spin averaged Altarelli-Parisi splitting function \cite{Altarelli:1977zs}. In \eqref{A04coll}, ``angular'' indicates the presence of angular terms due to spin correlations. The handling of these terms in the context of the antenna subtraction formalism is discussed in \cite{GRGG1,Glover:2010im}.
\subsubsection{Double-soft gluon limit}
\label{DoubleSoftFactor}
We 
 consider the leading-color antenna $A^0_4$ in the limit
  where the momenta $p_3$ and $p_4$ of the two gluons become simultaneously soft. This limit is 
  defined by rescaling the gluon momenta by an overall factor $\lambda$:
\begin{equation}
 p_3 \, \to \, \lambda p_3 \,, \quad p_4 \, \to \, \lambda p_4\, .
\end{equation}
In the limit $\lambda \to 0$ the quark-antiquark antenna 
$A^0_4$  behaves as
\begin{equation}
 A^0_4\! \left( 1_Q, 3_g , 4_g , 2_{\bar{Q}} \right) \,  \longrightarrow  \,  S( 1_Q, 3_g, 4_g, 2_{\bar{Q}} )   / \lambda^4  + \ldots \,,
\end{equation}
where the ellipses denote less singular contributions. The dominant singular term of
 $\cO ( 1 / 
\lambda^4 )$, the double-soft gluon function $S$, is given by
\mycomment{\begin{eqnarray}
 \lefteqn{  S( 1_Q, 3_g, 4_g, 2_{\bar{Q}} ) = 
\frac{ 2 s_{12}^2 }{ s_{13} s_{134} s_{24} s_{234} } } \quad 
\nonumber \\[0.1cm]
& & {} + \frac{ 2 s_{12} }{ s_{34} } \left( \frac{1}{ s_{13} s_{24} } + \frac{ 1 
}{ s_{13} s_{234} } + \frac{ 1 }{ s_{24} s_{ 134 } } - \frac{ 4 }{ s_{134} 
s_{234} } \right) + \frac{ 2 \left( 1 - \epsilon \right) }{ s_{34}^2 } \left( 
\frac{ s_{13} }{ s_{134} } + \frac{ s_{24} }{ s_{234} } - 1  \right)^2 
\nonumber \\[0.1cm]
& & {} + 4 m^2_Q \bigg[ -\frac{ s_{12} }{ s_{13} s_{24} } \left( \frac{1}{ s_{234} 
s_{24} } +  \frac{ 1 }{ s_{13} s_{134} } \right) + \frac{ 1 }{ s_{34} } \bigg(  
\frac{1}{s_{134}^2} + \frac{1}{s_{234}^2} - \frac{ s_{23} }{ s_{13} s_{134} 
s_{24} } - \frac{ s_{13} }{ s_{134} s_{234} s_{24} } 
\nonumber \\[0.1cm]
& & {} - \frac{ s_{14}}{ s_{13} s_{234} s_{24} } - \frac{ s_{24} }{s_{13} 
s_{134} s_{234}} \bigg) \bigg]
 + 4 m^4_Q \left[ \frac{1}{ s_{13}^2 s_{134}^2} + \frac{1}{ s_{13}^2 s_{24}^2 } + 
\frac{1}{ s_{234}^2 s_{24}^2} \right].
\label{DoubleSoftFunction}
\end{eqnarray}
}
\begin{eqnarray}
 \lefteqn{  S( 1_Q, 3_g, 4_g, 2_{\bar{Q}} ) = 
\frac{ 2 s_{12}^2 }{ s_{13} \left( s_{13} + s_{14} \right) s_{24} \left( s_{23} + s_{24} \right) } 
+ \frac{ 2 \left( 1 - \epsilon \right) }{ s_{34}^2 } \left( 
\frac{ s_{13} }{ \left( s_{13} + s_{14} \right) } + \frac{ s_{24} }{ \left( s_{23} + s_{24} \right) } - 1  \right)^2
} \quad 
\nonumber \\[0.1cm]
& & {} + \frac{ 2 s_{12} }{ s_{34} } \left( \frac{1}{ s_{13} s_{24} } + \frac{ 1 
}{ s_{13} \left( s_{23} + s_{24} \right) } + \frac{ 1 }{ s_{24} \left( s_{13} + s_{14} \right) } - \frac{ 4 }{ \left( s_{13} + s_{14} \right) 
\left( s_{23} + s_{24} \right) } \right)
\nonumber \\[0.1cm]
& & {}  
% \nonumber \\[0.1cm]
%
% & & {} 
+ 4 m^2_Q \bigg[ -\frac{ s_{12} }{ s_{13} s_{24} } \left( \frac{1}{ \left( s_{23} + s_{24} \right) 
s_{24} } +  \frac{ 1 }{ s_{13} \left( s_{13} + s_{14} \right) } \right) 
+ \frac{ 1 }{ s_{34} } \bigg(  
\frac{1}{\left( s_{13} + s_{14} \right)^2} + \frac{1}{\left( s_{23} + s_{24} \right)^2} \quad
\nonumber \\[0.1cm]
& & {}  - \frac{ s_{23} }{ s_{13} \left( s_{13} + s_{14} \right) 
s_{24} } - \frac{ s_{14}}{ s_{13} \left( s_{23} + s_{24} \right) s_{24} }
- \frac{ s_{24} }{s_{13} 
\left( s_{13} + s_{14} \right) \left( s_{23} + s_{24} \right)} 
\nonumber \\[0.1cm]
& & {}  - \frac{ s_{13} }{ \left( s_{13} + s_{14} \right) \left( s_{23} + s_{24} \right) s_{24} }\bigg) \bigg]
+ 4 m^4_Q \left[ 
  \frac{1}{ s_{13}^2 \left( s_{13} + s_{14} \right)^2} 
+ \frac{1}{ s_{13}^2 s_{24}^2 } 
+ \frac{1}{ \left( s_{23} + s_{24} \right)^2 s_{24}^2} 
\right]. \quad
\label{DoubleSoftFunction}
\end{eqnarray}
In the massless limit $S$ coincides with the result known from the literature (see e.g.~\cite{GRGG1}).

The subleading-color antenna $\tilde{ A }_4^0 \! \left( 1_Q , 3_g , 4_g , 
2_{\bar{Q}}  \right)$ is not color-ordered (and hence symmetric with respect to the 
interchange of the two gluon momenta $p_4$ and $p_3$).  Therefore it shows
QED-like factorization in the  double-soft limit:
\begin{equation}
 \tilde{ A }^0_4 \! \left( 1_Q, 3_g, 4_g, 2_{\bar{Q}} \right) \, \stackrel{ 3_g 
, 4_g \to 0}{ \longrightarrow } \, S(1_Q,3_g,2_{\bar{Q}}) \, S(1_Q,4_g,2_{\bar{Q}}) \,,
\end{equation}
where the single-soft eikonal function is given in \eqref{SingleEikonal}.
%

%
% version 18.9.2013 WB
%  version 15.9.2013 OD
%----------------------------------------------
%	Technical Details
%----------------------------------------------
%
\section{Integrated antenna functions 
\texorpdfstring{\boldmath $\mathcal{A}_{4}^{0}$}{ A04 } and 
\texorpdfstring{\boldmath $\tilde{\mathcal{A}}_{4}^{0}$ }{ A04t } }
\label{sec::Integrated_antenna}
In section \ref{sec::antenna_fct} we outlined the construction of $
d\sigma_{NNLO}^{S,\, Q\bar{Q}gg}$ whose integrated counter-part 
involves the integrated antenna functions $\mathcal{A}_{4}^{0}$ and $\tilde{\mathcal{A}}_{4}^{0}$.
  In this section  we compute these functions analytically. 
 They are obtained from $A_{4}^{0}$ and $\tilde{A}_{4}^{0}$ as follows:
\begin{eqnarray}
\cA^0_{4} \! \left( \epsilon; s_{1234}, m_Q \right)  & = & 
\frac{1}{ ( C( \epsilon ) )^2 }
\int\! d\Phi_{X_{Q\bar{Q}gg}} 
\, A_{4}^{0}\!\left(1_{Q},\,3_{g},\,4_{g},\,2_{\bar{Q}}\right),
\\[1ex]
\tilde{\cA}^0_{4} \! \left( \epsilon; s_{1234}, m_Q \right) & = & 
%\left(8\pi^{2}(4\pi)^{-\epsilon}e^{\epsilon\gamma_{E}}\right)^{2}
\frac{1}{ ( C( \epsilon ) )^2 }
\int\! d\Phi_{X_{Q\bar{Q}gg}} \, \tilde{A}_{4}^{0}\!\left(1_{Q},\,3_{g},\,4_{g},\,2_{\bar{Q}}\right).
\end{eqnarray}
% \left(8\pi^{2}(4\pi)^{-\epsilon}e^{\epsilon\gamma_{E}}\right)^{2}
Here the antenna phase-space measure $d\Phi_{X_{Q\bar{Q}gg}}$ is defined by 
\begin{equation}
d\Phi^{(D)}_{4}(p_{1},\, p_{2},\, p_{3},\, p_{4};\, q)=P_{2}(q^{2},\, m^2_Q)\, d\Phi_{X_{Q\bar{Q}gg}},
\end{equation}
where $d\Phi^{(D)}_{4}$ is the four-particle phase-space measure defined in eq.~\eqref{gm_3} and $P_{2}$ is the integrated two-particle phase space,
\begin{equation}
P_{2}(q^{2},\, m_Q^2 )=2^{-3+2\epsilon}\pi^{-1+\epsilon}\frac{\Gamma(1-\epsilon)}{\Gamma(2-2\epsilon)}\left(\frac{\mu^{2}}{q^{2}}\right)^{\epsilon}\left(1-\frac{4 m_Q^{2}}{q^{2}}\right)^{\frac{1}{2}-\epsilon}.
\end{equation}
First, we rewrite the four-particle phase
space measure according to 
\begin{eqnarray}
d\Phi^{(D)}_{4}(p_{1},\, p_{2},\, p_{3},\, p_{4};\, q) & = & \frac{\mu^{12-3D}}{i^{4}(2\pi)^{3D}}\delta^{(D)}\left(q-\sum_{i=1}^{4}p_{i}\right)\prod_{i=1}^{4}\frac{d^{D}p_{i}}{D_{i}},
\end{eqnarray}
 with the cut-propagators \cite{Cut, Anastasiou:2002yz}
\begin{eqnarray}
\frac{1}{D_{i}} & = & 2\pi i\delta^{+}(p_{i}^{2}-m_Q^{2})=\frac{1}{p_{i}^{2}-m_Q^{2}+i0}-\frac{1}{p_{i}^{2}-m_Q^{2}-i0}  \qquad \textrm{for }i=1,\,2,\\
\frac{1}{D_{i}} & = & 2\pi i\delta^{+}(p_{i}^{2})=\frac{1}{p_{i}^{2}+i0}-\frac{1}{p_{i}^{2}-i0} \qquad \textrm{for }i=3,\,4.
\end{eqnarray}
 We introduce six further propagators $D_{5}, \ldots , D_{10}$ such that we can express all scalar products $s_{ij},\, s_{ijk}$ 
in the integrands $A_{4}^{0},\,\tilde{A}_{4}^{0}$ by the functions $D_{1},\ldots, D_{10}$. The set of these ten propagators is linearly dependent. However,
 each term in the two integrands can be expressed by at most nine or less propagators. 
The terms can be treated as integrands of cut-integrals, corresponding to four-particle cuts through 
three-loop propagator-type Feynman graphs. We distribute the terms to appropriate topologies, each given 
by nine independent propagators, and apply integration-by-parts reduction \cite{CheTka}, using the 
implementation {\tt FIRE} \cite{Smi} of the Laporta algorithm \cite{Lap}. 

As a result of this reduction we can express the integrated antenna functions 
 $\cA_{4}^{0}$ and $\tilde{\cA}_{4}^{0}$ by the following 15 master integrals:
\begin{eqnarray}
T_1 (q^2 , m_Q^2, \epsilon ) & = & 
\parbox{0.15\linewidth}{
\resizebox{\linewidth}{!}{
\fcolorbox{white}{white}{
  \begin{picture}(194,162) (63,-31)
    \SetWidth{4.0}
    \SetColor{Black}
    \Arc(160,50)(64,180,540)
    \SetWidth{1.0}
    \Line[double,sep=4](96,50)(64,50)
    \Line[double,sep=4](224,50)(256,50)
    \Arc[clock](160,-23.333)(97.333,131.112,48.888)
    \Arc(160,123.333)(97.333,-131.112,-48.888)
    \Line[dash,dashsize=4.6](160,130)(160,-30)
  \end{picture}
}}}
= \int \! d\Phi^{(D)}_4(p_1, p_2, p_3, p_4; q)\,,
\label{Master1} \\[0.2cm]
T_2 (q^2 , m_Q^2, \epsilon ) & = & \raisebox{2ex}{\makebox[5ex]{$s_{13}$}}
\hspace{ - 2ex }
\parbox{0.15\linewidth}{
\resizebox{\linewidth}{!}{
  \begin{picture}(194,162) (63,-31)
    \SetWidth{4.0}
    \SetColor{Black}
    \Arc(160,50)(64,180,540)
    \SetWidth{1.0}
    \Line[double,sep=4](96,50)(64,50)
    \Line[double,sep=4](224,50)(256,50)
    \Arc[clock](160,-23.333)(97.333,131.112,48.888)
    \Arc(160,123.333)(97.333,-131.112,-48.888)
    \Line[dash,dashsize=4.6](160,130)(160,-30)
  \end{picture}
}}
\mycomment{ \parbox{0.15\linewidth}{
\resizebox{\linewidth}{!}{
\fcolorbox{white}{white}{
  \begin{picture}(208,162) (49,-31)
    \SetWidth{4.0}
    \SetColor{Black}
    \Arc(160,50)(64,180,540)
    \SetWidth{1.0}
    \Line[double,sep=4](96,50)(64,50)
    \Line[double,sep=4](224,50)(256,50)
    \Arc[clock](160,-23.333)(97.333,131.112,48.888)
    \Arc(160,123.333)(97.333,-131.112,-48.888)
    \Line[dash,dashsize=4.6](160,130)(160,-30)
    \Text(64,66)[]{\Huge{\Black{$s_{13}$}}}
  \end{picture}
}}} }
= \int \! d\Phi^{(D)}_4(p_1, p_2, p_3, p_4; q) \: s_{13}\,, 
\label{Master2} \\[0.2cm]
T_3 (q^2 , m_Q^2, \epsilon ) & = & \raisebox{2ex}{\makebox[5ex]{$s_{134}$}}
\hspace{ - 2ex }
\parbox{0.15\linewidth}{
\resizebox{\linewidth}{!}{
  \begin{picture}(194,162) (63,-31)
    \SetWidth{4.0}
    \SetColor{Black}
    \Arc(160,50)(64,180,540)
    \SetWidth{1.0}
    \Line[double,sep=4](96,50)(64,50)
    \Line[double,sep=4](224,50)(256,50)
    \Arc[clock](160,-23.333)(97.333,131.112,48.888)
    \Arc(160,123.333)(97.333,-131.112,-48.888)
    \Line[dash,dashsize=4.6](160,130)(160,-30)
  \end{picture}
}}
\mycomment{ \parbox{0.15\linewidth}{
\resizebox{\linewidth}{!}{
\fcolorbox{white}{white}{
  \begin{picture}(208,162) (49,-31)
    \SetWidth{4.0}
    \SetColor{Black}
    \Arc(160,50)(64,180,540)
    \SetWidth{1.0}
    \Line(96,50)(64,50)
    \Line(224,50)(256,50)
    \Arc[clock](160,-23.333)(97.333,131.112,48.888)
    \Arc(160,123.333)(97.333,-131.112,-48.888)
    \Line[dash,dashsize=4.6](160,130)(160,-30)
    \Text(64,66)[]{\Huge{\Black{$s_{134}$}}}
  \end{picture}
}}} }
= \int \! d\Phi^{(D)}_4(p_1, p_2, p_3, p_4; q) \:
s_{134}\,, 
\label{Master3} \\[0.2cm]
T_4 (q^2 , m_Q^2, \epsilon ) & = & 
\parbox{0.15\linewidth}{
\resizebox{\linewidth}{!}{
\fcolorbox{white}{white}{
  \begin{picture}(194,162) (63,-31)
    \SetWidth{4.0}
    \SetColor{Black}
    \Arc(160,50)(64,180,540)
    \SetWidth{1.0}
    \Line[double,sep=4](96,50)(64,50)
    \Line[double,sep=4](224,50)(256,50)
    \Arc[clock](140.343,-23.835)(111.581,108.466,41.431)
    \Arc(185.901,145.585)(102.898,-141.834,-68.268)
    \Line[dash,dashsize=4.6](160,130)(160,-30)
  \end{picture}
}}}
= \int \! d\Phi^{(D)}_4(p_1, p_2, p_3, p_4; q) \:
\frac{1}{s_{134}}\,, 
\label{Master4} \\[0.2cm]
T_5 (q^2 , m_Q^2, \epsilon ) & = & 
\parbox{0.15\linewidth}{
\resizebox{\linewidth}{!}{
\fcolorbox{white}{white}{
  \begin{picture}(194,162) (63,-31)
    \SetWidth{4.0}
    \SetColor{Black}
    \Arc(160,50)(64,180,540)
    \SetWidth{1.0}
    \Line[double,sep=4](96,50)(64,50)
    \Line[double,sep=4](224,50)(256,50)
    \Arc[clock](119.345,-19.876)(102.881,98.015,21.602)
    \Arc(200.655,119.876)(102.881,-158.398,-81.985)
    \Line[dash,dashsize=4.6](160,130)(160,-30)
  \end{picture}
}}}
 = \int \! d\Phi^{(D)}_4(p_1, p_2, p_3, p_4; q) \:
\frac{1}{s_{134} s_{234}} \,,
\label{Master5} \\[0.2cm]
T_6 (q^2 , m_Q^2, \epsilon ) & = & 
\parbox{0.15\linewidth}{
\resizebox{\linewidth}{!}{
  \begin{picture}(194,162) (63,-31)
    \SetWidth{4.0}
    \SetColor{Black}
    \Arc(160,50)(64,180,540)
    \SetWidth{1.0}
    \Line[double,sep=4](96,50)(64,50)
    \Line[double,sep=4](224,50)(256,50)
    \Line[dash,dashsize=4.6](160,130)(160,-30)
    \Line(97,50)(217,18)
    \Line(117,99)(217,18)
  \end{picture}
}}
= \int \! d\Phi^{(D)}_4(p_1, p_2, p_3, p_4; q)\: \frac{1}{s_{13} s_{234}} \, ,
\label{masters::6} \\[0.2cm]
T_7 (q^2 , m_Q^2, \epsilon ) & = & 
\parbox{0.15\linewidth}{
\resizebox{\linewidth}{!}{
  \begin{picture}(194,162) (63,-31)
    \SetWidth{4.0}
    \SetColor{Black}
    \Arc(160,50)(64,180,540)
    \SetWidth{1.0}
    \Line[double,sep=4](96,50)(64,50)
    \Line[double,sep=4](224,50)(256,50)
    \Line[dash,dashsize=4.6](160,130)(160,-30)
    \Line(103,79)(217,18)
    \Line(131,108)(217,18)
  \end{picture}
}}
= \int \! d\Phi^{(D)}_4(p_1, p_2, p_3, p_4; q) \: \frac{1}{s_{13} s_{134} s_{234}} \,, 
\label{masters::7} \\[0.2cm]
T_8 (q^2 , m_Q^2, \epsilon ) & = & \raisebox{2ex}{\makebox[5ex]{$s_{24}$}}
\hspace{ - 2ex }
\parbox{0.15\linewidth}{
\resizebox{\linewidth}{!}{
  \begin{picture}(194,162) (63,-31)
    \SetWidth{4.0}
    \SetColor{Black}
    \Arc(160,50)(64,180,540)
    \SetWidth{1.0}
    \Line[double,sep=4](96,50)(64,50)
    \Line[double,sep=4](224,50)(256,50)
    \Line[dash,dashsize=4.6](160,130)(160,-30)
    \Line(103,79)(217,18)
    \Line(131,108)(217,18)
  \end{picture}
}}
= \int \! d\Phi^{(D)}_4(p_1, p_2, p_3, p_4; q) \: \frac{ s_{24} }{ s_{13} s_{134} s_{234} } \,,
\label{masters::8} 
\\[0.2cm]
% \end{eqnarray}
%
%
%\begin{eqnarray}
T_9 (q^2 , m_Q^2, \epsilon ) & = & 
\parbox{0.15\linewidth}{
\resizebox{\linewidth}{!}{
  \begin{picture}(194,162) (63,-31)
    \SetWidth{4.0}
    \SetColor{Black}
    \Arc(160,50)(64,180,540)
    \SetWidth{1.0}
    \Line[double,sep=4](96,50)(64,50)
    \Line[double,sep=4](224,50)(256,50)
    \Line[dash,dashsize=4.6](160,130)(160,-30)
    \Line(97,50)(203,99)
    \Line(117,99)(154,82)
    \Line(167,76)(223,50)
  \end{picture}
}}
= \int \! d\Phi^{(D)}_4(p_1, p_2, p_3, p_4; q)\: \frac{1}{s_{13} s_{14}} \, ,
\label{masters::9} \\[0.2cm]
T_{10} (q^2 , m_Q^2, \epsilon ) & = & \raisebox{2ex}{\makebox[5ex]{$s_{23}$}}
\hspace{ - 2ex }
\parbox{0.15\linewidth}{
\resizebox{\linewidth}{!}{
  \begin{picture}(194,162) (63,-31)
    \SetWidth{4.0}
    \SetColor{Black}
    \Arc(160,50)(64,180,540)
    \SetWidth{1.0}
    \Line[double,sep=4](96,50)(64,50)
    \Line[double,sep=4](224,50)(256,50)
    \Line[dash,dashsize=4.6](160,130)(160,-30)
    \Line(97,50)(203,99)
    \Line(117,99)(154,82)
    \Line(167,76)(223,50)
  \end{picture}
}}
= \int \! d\Phi^{(D)}_4(p_1, p_2, p_3, p_4; q) \: \frac{s_{23}}{s_{13} s_{14}} \,, 
\label{masters::10} \\[0.2cm]
T_{11} (q^2 , m_Q^2, \epsilon ) & = & \raisebox{2ex}{\makebox[5ex]{$s_{134}$}}
\hspace{ - 2ex }
\parbox{0.15\linewidth}{
\resizebox{\linewidth}{!}{
  \begin{picture}(194,162) (63,-31)
    \SetWidth{4.0}
    \SetColor{Black}
    \Arc(160,50)(64,180,540)
    \SetWidth{1.0}
    \Line[double,sep=4](96,50)(64,50)
    \Line[double,sep=4](224,50)(256,50)
    \Line[dash,dashsize=4.6](160,130)(160,-30)
    \Line(97,50)(203,99)
    \Line(117,99)(154,82)
    \Line(167,76)(223,50)
  \end{picture}
}}
= \int \! d\Phi^{(D)}_4(p_1, p_2, p_3, p_4; q) \: \frac{ s_{134} }{s_{13} s_{14}} \,,
\label{masters::11}  \\[0.2cm]
T_{12} (q^2 , m_Q^2, \epsilon ) & = &
\parbox{0.15\linewidth}{
\resizebox{\linewidth}{!}{
  \begin{picture}(194,162) (63,-31)
    \SetWidth{4.0}
    \SetColor{Black}
    \Arc(160,50)(64,180,540)
    \SetWidth{1.0}
    \Line[double,sep=4](96,50)(64,50)
    \Line[double,sep=4](224,50)(256,50)
    \Line[dash,dashsize=4.6](160,130)(160,-30)
    \Line(115,5)(205,95)
    \Line(115,95)(155,55)
    \Line(165,45)(205,5)
  \end{picture}
}}
= \int \! d\Phi^{(D)}_4(p_1, p_2, p_3, p_4; q) \: \frac{ 1 }{ s_{13} s_{14} s_{23} s_{24} } \,,
\label{masters::12}  \\[0.2cm]
T_{13} (q^2 , m_Q^2, \epsilon ) & = & \raisebox{2ex}{\makebox[5ex]{$s_{134}$}}
\hspace{ - 2ex }
\parbox{0.15\linewidth}{
\resizebox{\linewidth}{!}{
  \begin{picture}(194,162) (63,-31)
    \SetWidth{4.0}
    \SetColor{Black}
    \Arc(160,50)(64,180,540)
    \SetWidth{1.0}
    \Line[double,sep=4](96,50)(64,50)
    \Line[double,sep=4](224,50)(256,50)
    \Line[dash,dashsize=4.6](160,130)(160,-30)
    \Line(115,5)(205,95)
    \Line(115,95)(155,55)
    \Line(165,45)(205,5)
  \end{picture}
}}
= \int \! d\Phi^{(D)}_4(p_1, p_2, p_3, p_4; q) \: \frac{ s_{134} }{ s_{13} s_{14} s_{23} s_{24} } \,,
\label{masters::13} \\[0.2cm]
T_{14} (q^2 , m_Q^2, \epsilon ) & = & 
\parbox{0.15\linewidth}{
\resizebox{\linewidth}{!}{
  \begin{picture}(194,162) (63,-31)
    \SetWidth{4.0}
    \SetColor{Black}
    \Arc(160,50)(64,180,540)
    \SetWidth{1.0}
    \Line[double,sep=4](96,50)(64,50)
    \Line[double,sep=4](224,50)(256,50)
    \Line[dash,dashsize=4.6](160,130)(160,-30)
    \Line(101,73)(219,27)
    \Line(128,105)(157,55)
    \Line(163,45)(192,-5)
  \end{picture}
}}
= \int \! d\Phi^{(D)}_4(p_1, p_2, p_3, p_4; q)\: \frac{1}{s_{13} s_{23} s_{134} 
s_{234} } \, ,
\label{masters::4} \\[0.2cm]
T_{15} (q^2 , m_Q^2, \epsilon ) & = & \raisebox{2ex}{\makebox[5ex]{$s_{14}$}}
\hspace{ - 2ex }
\parbox{0.15\linewidth}{
\resizebox{\linewidth}{!}{
  \begin{picture}(194,162) (63,-31)
    \SetWidth{4.0}
    \SetColor{Black}
    \Arc(160,50)(64,180,540)
    \SetWidth{1.0}
    \Line[double,sep=4](96,50)(64,50)
    \Line[double,sep=4](224,50)(256,50)
    \Line[dash,dashsize=4.6](160,130)(160,-30)
    \Line(101,73)(219,27)
    \Line(128,105)(157,55)
    \Line(163,45)(192,-5)
  \end{picture}
}}
= \int \! d\Phi^{(D)}_4(p_1, p_2, p_3, p_4; q) \: \frac{ s_{14} }{s_{13} s_{23} 
s_{134} s_{234} } \,.
\label{masters::15}
\end{eqnarray}
In these diagrammatic representations bold (thin) lines refer to massive 
(massless) propagators. The invariants to the left of the cut-diagrams denote 
numerators of the integrand. The dashed line stands for the four-particle cut 
and the edges crossed by the dashed lines correspond to the cut-propagators 
$D_{1},\ldots,D_{4}$. The integrated leading-color antenna $\cA^0_4$ can 
be expressed in terms of the master integrals $T_1, \ldots, T_8$, whereas the 
computation of the subleadig-color antenna $\tilde{\cA}^0_4$ involves the entire 
set of the above master integrals. 

Analytic results for the integrals $T_{1},\ldots, T_{5}$ were already given in \cite{Bernreuther:2011jt} in 
terms of harmonic polylogarithms \cite{RemVer} of argument
\begin{equation}
y = \frac{ 1 - \sqrt{ 1 - \frac{4 m^2_Q}{q^2} } }{ 1 + \sqrt{ 1 - \frac{ 4 m_Q^2 }{q^2} } }\,.
\end{equation}
The expansion of these expressions near $D = 4$ was given  in \cite{Bernreuther:2011jt} to 
$\cO( \epsilon^2 )$ in the case of $T_1$, $T_2$ and $T_3$ and to order $\cO( 
\epsilon )$ in the case of $T_4$ and $T_5$. Here,
  the $\cO( \epsilon^3 )$-terms of $T_1$, $T_2$ and $T_3$ are also required.
 We computed them from the results provided in \cite{Bernreuther:2011jt} by using the computer programs
 {\tt HypExp} \cite{HubMai} and {\tt HYPERDIRE} \cite{Bytev:2011ks}.

We compute the remaining ten master integrals $T_{6},\ldots,T_{15}$ with
  the same techniques that were used
  in the computation of $T_4$ and $T_5$ in \cite{Bernreuther:2011jt}. 
 Applying the method of  differential equations 
  \cite{Kot, Rem, Gehrmann:1999as, ArgMas}, 
 we derive inhomogeneous first order differential equations in the variables $q^2 = s_{1234} + 2m_Q^2 $ and $y$ for each master integral. 
In each case, the inhomogeneous part of such a differential equation for a master integral $T_i$ is expressed as linear combination of 
some of the other master integrals. We analyze
 these equations order by order in $\epsilon$. 
 If the resulting system of equations decouples and if the inhomogeneous part can be expressed by known results 
 in terms of iterated integrals, we can derive the solution of the corresponding decoupled
  equation.
  This involves the $y$-integration of the inhomogeneous part. 
 Following this procedure we make use of our results for $T_{1},\ldots , T_{5}$. 
 These results allow us to begin with those equations where the inhomogeneous part only involves these integrals. Starting from there, we then proceed with the more complicated equations. Using partial integration, partial fraction decomposition and the package {\tt HPL} \cite{Mai} for the integrations, we obtain analytical results for all the above  master integrals in terms of harmonic polylogarithms. The integration constants are fixed by using that the above master integrals vanish at threshold. Explicit expressions for the master integrals can be obtained from the authors upon request.
\mycomment{bottom up approach, Laurant expansion. Up to required orders.}

We then obtain as our main result the integrated antenna functions  $\mathcal{A}_{4}^{0}$ and $\tilde{\mathcal{A}}_{4}^{0}$ in terms of harmonic polylogarithms:
\begin{eqnarray}
\lefteqn{
\cA^0_{4} \! \left( \epsilon; s_{1234}, m_Q \right) 
% & = &
=
\left( s_{1234} + 2 m_Q^2 \right)^{- 2 \epsilon } \Bigg\{ \frac{1}{ \epsilon^3 }
\Bigg[
\frac{1}{2} - \left( \frac{1}{2} - \frac{1}{2 (1-y)} - \frac{1}{2 (1+y)} 
\right) \text{H} ( 0 ; y )
\Bigg]
}
\qquad
\nonumber \\
& & {} + \frac{1}{ \epsilon^2 } \Bigg[
\left( 
  \frac{19}{12} 
- \frac{5}{6 (1-y)} 
+ \frac{13}{6 (1+y)}
- \frac{ 1 + 2 y}{2 \left( 1 + 4 y + y^2 \right) } 
\right) 
\text{H} ( 0 ; y )
+ 4 \text{H} ( 1 ; y )
\nonumber \\
& & {} 
+ \left(
  1
- \frac{1}{1-y}
- \frac{1}{1+y}
\right) 
\left( 
\text{H} ( 1,0 ; y ) - 4 \text{H} ( 0,1 ; y ) 
- 3 \text{H} ( -1,0 ; y ) + \frac{7}{2} \zeta(2)
\right)
\nonumber \\
& & {} 
+ \frac{11}{3} 
+ \frac{3 y}{1+4 y+y^2} 
\Bigg]
% \nonumber \\
%
% & & {} 
+ \frac{1}{ \epsilon } \Bigg[
\left(
  21  
- \frac{ 21 }{ 1 - y }
- \frac{ 21 }{ 1 + y } 
\right)
\zeta(2) \, \text{H} ( -1 ; y )
\nonumber \\
& & {}
+ \left( 
  \frac{361}{18} + 4 \zeta (2)
+ \frac{ \zeta (2) }{2 (1-y)^2}
- \frac{1201 + 288 \zeta (2) }{ 72 ( 1 - y ) }
+ \frac{ \zeta (2) }{2 (1+y)^2}
\right.
\nonumber \\
& & {} 
+ \frac{95 - 288 \zeta (2) }{72 (1+y)}
- \frac{2 + 7 y}{\left(1+4 y+y^2\right)^2}
+ \frac{79+248 y}{12 \left(1+4 y+y^2\right)} \,
\bigg) \,
\text{H} ( 0 ; y )
\nonumber \\
& & {} + \left(
  \frac{88}{3} - 7 \zeta(2) 
+ \frac{7 \zeta(2) }{1-y}
+ \frac{7 \zeta(2) }{1+y}
+ \frac{24 y}{1+4 y+y^2}
\right) 
\text{H} ( 1 ; y )
\nonumber \\
& & { } + \left(
  \frac{38}{3}
- \frac{20}{3 (1-y)}
+ \frac{52}{3 (1+y)}
- \frac{4 (1+2 y)}{1+4 y+y^2}
\right) 
\text{H} ( 0, 1 ; y )
\nonumber \\
& & {} - \left(
  \frac{10}{3} 
+ \frac{25}{6 (1-y)}
- \frac{83}{6 (1+y)}
+ \frac{3 (1+2 y)}{1+4 y+y^2} 
\right) 
\text{H} ( -1,0 ; y ) 
\nonumber \\
& & {} - \left(
  \frac{7}{6}
+ \frac{5}{6 (1-y)}
+ \frac{5}{6 (1+y)}
\right) 
\text{H} ( 0,0 ; y )
+ 32\, \text{H} ( 1,1 ; y )
\nonumber \\
& & { } + \left(
  \frac{53}{3}
+  \frac{5}{6 (1-y)}
-  \frac{31}{6 (1+y)}
+  \frac{1+2 y}{1+4 y+y^2}
\right) 
\text{H} ( 1,0 ; y )
\nonumber \\
& & {} + \left(
  11
- \frac{1}{(1-y)^2}
- \frac{11}{1-y}
- \frac{1}{(1+y)^2}
- \frac{11}{1+y}
\right) 
\text{H} ( 0,-1,0 ; y )
\nonumber \\
& & { } - \left(
  19
- \frac{1}{(1-y)^2}
- \frac{19}{1-y}
- \frac{1}{(1+y)^2}
- \frac{19}{1+y}
\right) 
\text{H} ( 0,1,0 ; y )
\nonumber \\
& & { } + \left(
  3
+ \frac{1}{(1-y)^2}
- \frac{3}{1-y}
+ \frac{1}{(1+y)^2}
- \frac{3}{1+y}
\right) 
\text{H} ( 0,0,0 ; y )
\nonumber \\
& & { } - \left(
  24
- \frac{24}{1-y}
- \frac{24}{1+y}
\right) 
\text{H} ( -1,0,1 ; y )
+\left(
  8
- \frac{8}{1-y}
- \frac{8}{1+y}
\right) 
\text{H} ( 1,0,1 ; y )
\nonumber \\
& & {} - \left(
  32
- \frac{32}{1-y}
- \frac{32}{1+y}
\right) 
\text{H} ( 0,1,1 ; y )
-\left(
  18
- \frac{18}{1-y}
- \frac{18}{1+y}
\right) 
\text{H}(-1,-1,0; y )
\nonumber \\
& & {} + \left(
  6
- \frac{6}{1-y}
- \frac{6}{1+y}
\right) 
\text{H} ( -1,1,0 ; y )
+ \left(
  6
- \frac{6}{1-y}
- \frac{6}{1+y}
\right) 
\text{H} ( 1,-1,0 ; y )
\nonumber \\
& & {} - \left(
  2
- \frac{2}{1-y}
- \frac{2}{1+y}
\right) 
\text{H} ( 1,1,0 ; y )
+ \frac{1595}{72}
+ \frac{6 (1+4 y)}{\left(1+4 y+y^2\right)^2}
- \frac{2 (3 - 13 y)}{1+4 y+y^2}
\nonumber \\
& & { } - \left( 
  \frac{85}{6}
- \frac{65}{12 (1-y)}
+ \frac{187}{12 (1+y)}
- \frac{7 (1+2 y)}{2 \left(1+4 y+y^2\right)}
\right) \zeta ( 2 )
\nonumber \\
& & {}  + \left( 
  12
- \frac{12}{1-y}
- \frac{12}{1+y}
+ \frac{1}{2 (1-y)^2}
+ \frac{1}{2 (1+y)^2}
\right) \zeta(3)
\Bigg] + F(y) \Bigg\} \,,
\\[3ex]
%
% >>> A04t
%
\lefteqn{
\tilde{\cA}^0_{4} \! \left( \epsilon; s_{1234}, m_Q \right) 
% & = &
= 
\left( s_{1234} + 2 m_Q^2 \right)^{- 2 \epsilon } \Bigg\{ 
%
% >>> eps -2
%
\frac{1}{ \epsilon^2 }
\Bigg[
1 - \left( 2 - \frac{2}{1-y} - \frac{2}{1+y} \right) \text{H} ( 0 ; y ) 
}
\qquad
\nonumber \\
& & {} + \left( 
  2 
- \frac{2}{1-y}
- \frac{2}{1+y} 
+ \frac{2}{(1-y)^2} 
+ \frac{2}{(1+y)^2} 
\right) \text{H}( 0,0 ; y ) 
\Bigg]
\nonumber \\
%
% >>> eps -1 
%
& & {} + \frac{1}{\epsilon} \Bigg[
8 \, \text{H} ( 1 ; y ) - \bigg( 
  1 
- \frac{ 25 }{6 (1-y)}
- \frac{ 31 }{2 (1+y)}
+ \frac{29 (1+2 y)}{3 \left(1+4 y+y^2\right)}
\nonumber \\
& & {} + \left( 
  12
- \frac{12}{1-y}
- \frac{12}{1+y} 
+ \frac{12}{(1-y)^2}
+ \frac{12}{(1+y)^2}
\right) \zeta ( 2 ) 
\bigg)
\text{H} ( 0 ; y )
\nonumber \\
& & {} + \left(
  4
- \frac{4}{1-y} 
- \frac{4}{1+y}
\right) 
\text{H} ( 1,0 ; y )
- \left(
  16
- \frac{16}{1-y}
- \frac{16}{1+y}
\right) 
\text{H} ( 0, 1 ; y )
\nonumber \\
& & {} + \left(
  4
+ \frac{14}{3 (1-y)}
- \frac{2}{1+y}
- \frac{8}{(1-y)^2}
+ \frac{ 4 + 32 y }{3 \left(1+4 y+y^2\right)}
\right) 
\text{H} ( 0,0 ; y )
\nonumber \\
& & {} - \left(
  16
- \frac{56}{3 (1-y)}
- \frac{16}{1+y}
+ \frac{8 + 16 y }{3 \left(1+4 y+y^2\right)}
\right) 
\text{H} ( -1,0 ; y )
\nonumber \\
& & {} +\left(
  16
- \frac{16}{1-y}
- \frac{16}{1+y}
+ \frac{16}{(1-y)^2}
+ \frac{16}{(1+y)^2}
\right) 
\text{H} ( 0,0,1 ; y )
\nonumber \\
& & {} +\left(
  16
- \frac{16}{1-y}
- \frac{16}{1+y}
+ \frac{16}{(1-y)^2}
+ \frac{16}{(1+y)^2}
\right) 
\text{H} ( 0, -1, 0 ; y )
\nonumber \\
& & {} - \left(
  4
- \frac{4}{1-y}
- \frac{4}{1+y}
+ \frac{4}{(1-y)^2}
+ \frac{4}{(1+y)^2}
\right) 
\text{H} ( 0,1,0; y )
\nonumber \\
& & {} + \left(
  8 
- \frac{8}{1-y}
- \frac{8}{1+y}
+ \frac{8}{(1-y)^2}
+ \frac{8}{(1+y)^2}
\right)
\text{H} ( 1,0,0 ; y )
\nonumber \\
& & {} + \left(
  10
- \frac{32}{3 (1-y)}
- \frac{8}{1+y}
+ \frac{2}{(1-y)^2}
+ \frac{2}{(1+y)^2}
- \frac{4 + 8 y}{3 \left(1+4 y+y^2\right)}
\right) 
\text{H} ( 0,0,0 ; y )
\nonumber \\
& & {} + \frac{25}{2}
+ \frac{22  y }{ 1 +4 y +y^2}
+ \left( 
  12
+ \frac{32}{3 (-1+y)}
- \frac{12}{1+y}
- \frac{4 + 8 y}{3 \left(1+4 y+y^2\right)}
\right) \zeta ( 2 )
\nonumber \\
& & {} -\left( 
  8
- \frac{8}{1-y}
- \frac{8}{1+y} 
+ \frac{8}{(1-y)^2}
+ \frac{8}{(1+y)^2}
\right) \zeta ( 3 ) 
\Bigg]
+ {\tilde F}(y) \Bigg\} \, .
\end{eqnarray}
The finite remainders $F(y)$ and ${\tilde F}(y)$  of $\cO\! \left( \epsilon^0 \right)$ are also given analytically in terms of harmonic polylogarithms up to and including weight four. These expressions are quite long and can be obtained from the authors upon request.

%
% version 18.9.13 WB
% version 15.9.13, OD
%----------------------------------------------
%	Summary
%----------------------------------------------
%
\section{Summary and outlook}
\label{sec::conclusion}
We addressed the treatment of infrared singularities that arise
  in the 
computation of observables,
 in particular distributions, 
 for processes at NNLO QCD, where a heavy quark-pair 
is produced by an uncolored initial state. 
 In the framework of the antenna subtraction method, 
appropriate subtraction terms are construced in terms of
  universal antenna 
functions. We constructed the NNLO
antenna functions $A^0_{4},\,\tilde{A^0_{4}}$ for the 
real-radiation subtraction term that is required for the
 squared matrix element of the final state  that consists of
a  pair of heavy quarks and two gluons. We discussed   the 
singular limits  of this
 subtraction term and, as our main result, we computed the corresponding 
integrated antenna functions $\mathcal{A}^0_{4},\,\tilde{\mathcal{A}^0_{4}}$ 
analytically.  Our 
results  include also analytical expressions for a set of  master 
integrals, which we expect to be useful for other applications, too. 

The antenna functions $A^0_{4},\,\tilde{A^0_{4}}$ and their integrated versions 
$\mathcal{A}^0_{4},\,\tilde{\mathcal{A}^0_{4}}$ provide a further step towards the 
calculation of   $d\sigma_{\mss{NNLO}}$ for reactions of the type \eqref{intro:eq1} within the antenna framework.
  All building blocks for 
the real-radiation subtraction terms  and their integrated counter-parts are now 
available.  The missing piece, the 
real-virtual subtraction term  
$d\sigma_{\mss{NNLO}}^{T}$ (cf. \eqref{eq:cross_section_NNLO}), can be constructed from 
  the interference  of the tree-level and 1-loop matrix element
  for  $\gamma^*\to  Q {\bar Q} g$. Its integral over the 3-particle phase space
 can also be done in analytical form. We plan to present this result in a future publication
 \cite{WBOD}.

%
%----------------------------------------------
%	Acknowledgments
%----------------------------------------------
% 
\acknowledgments 
We thank Gabriel Abelof, Aude Gehrmann-De Ridder and Thomas Gehrmann for discussions. 
This work was supported  by Deutsche Forschungsgemeinschaft (DFG)
Sonderforschungsbereich/Trans\-regio 9 ``Computergest\"utzte Theoretische 
Teilchenphysik''. O.D. was supported in part by the Research 
Executive Agency (REA) of the European Union under the Grant Agreement number 
PITN-GA-2010-264564 (LHCPhenoNet). C.B. thanks Dirk Kreimer's group at Humboldt 
University for support. The figures were generated using Jaxodraw 
\cite{Binosi:2003yf}, based on Axodraw 
\cite{Vermaseren:1994je}.

\appendix
\section{Antenna functions 
\texorpdfstring{\boldmath $A_{4}^{0}$}{ A04 } and 
\texorpdfstring{\boldmath $\tilde{A}_{4}^{0}$ }{ A04t } }
\label{appendix::antennae}
For the numerical computation of \eqref{sbdifcrx} in $D = 4$ dimensions, only the four-dimensional parts of the antenna functions ${A}^0_4$ and $\tilde{A}^0_4$ are required. However, the integrated antenna functions $\cA^0_4$ and $\tilde{\cA}^0_4$, which we computed in sec.~\ref{sec::Integrated_antenna}, must be determined with $A^0_4$ and $\tilde{A}^0_4$ in $D$ dimensions. The four-dimensional parts of $A^0_4$ and $\tilde{A}^0_4$ read:
\begin{eqnarray}
A^0_4 \! \left( 1_Q, 3_g, 4_g, 2_{\bar{Q}} \right)
& = & \textstyle
\frac{ 1 } {
s_{1234} + 4 m_Q^2 } \Big( 
-\frac{2}{s_{13}}
-\frac{2}{s_{24}}
+\frac{3 s_{12}}{s_{134}^2}
+\frac{3 s_{12}}{s_{234}^2}
+\frac{5}{s_{134}}
+\frac{5}{s_{234}}
-\frac{6 s_{12}}{s_{13} s_{134}}
-\frac{6 s_{12}}{s_{234} s_{24}}
\nonumber \\
& & \textstyle {} 
-\frac{6 s_{12}}{s_{134} s_{24}}
-\frac{6 s_{12}}{s_{13} s_{234}}
+\frac{3 s_{23}}{s_{134}^2}
+\frac{3 s_{14}}{s_{234}^2}
-\frac{3 s_{23}}{s_{13} s_{134}}
-\frac{3 s_{14}}{s_{234} s_{24}}
+\frac{3 s_{13}}{s_{234}^2}
+\frac{3 s_{24}}{s_{134}^2}
\nonumber \\
& & \textstyle {} 
-\frac{4 s_{12}^2}{s_{13} s_{134} s_{234}}
-\frac{4 s_{12}^2}{s_{134} s_{234} s_{24}}
-\frac{7 s_{13}}{s_{134} s_{234}}
-\frac{7 s_{24}}{s_{134} s_{234}}
-\frac{3 s_{14}}{s_{13} s_{234}}
-\frac{3 s_{23}}{s_{134} s_{24}}
+\frac{6 s_{12}}{s_{13} s_{24}}
\nonumber \\
& & \textstyle {} 
+\frac{4 s_{12}^2}{s_{13} s_{134} s_{24}}
+\frac{4 s_{12}^2}{s_{13} s_{234} s_{24}}
+\frac{3 s_{13}}{s_{134} s_{24}}
+\frac{3 s_{24}}{s_{13} s_{234}}
+\frac{3 s_{14}}{s_{13} s_{24}}
+\frac{3 s_{23}}{s_{13} s_{24}}
+\frac{10 s_{12}}{s_{134} s_{234}}
\nonumber \\
& & \textstyle {} 
+\frac{3 s_{12} s_{23}}{s_{13} s_{134} s_{24}}
+\frac{3 s_{12} s_{14}}{s_{13} s_{234} s_{24}}
+\frac{s_{23}^2}{s_{13} s_{134} s_{24}}
+\frac{s_{14}^2}{s_{13} s_{234} s_{24}}
-\frac{s_{13}}{s_{234} s_{24}}
-\frac{s_{24}}{s_{13} s_{134}}
\nonumber \\
& & \textstyle {} 
+\frac{2 s_{12}^3}{s_{13} s_{134} s_{234} s_{24}}
+\frac{3 s_{12} s_{13}}{s_{134} s_{234} s_{24}}
+\frac{3 s_{12} s_{24}}{s_{13} s_{134} s_{234}}
-\frac{s_{13}^2}{s_{134} s_{234} s_{24}}
-\frac{s_{24}^2}{s_{13} s_{134} s_{234}}
+\frac{2 s_{12}}{s_{34}^2}
\nonumber \\
& & \textstyle {} 
-\frac{2 s_{13}}{s_{34}^2}
-\frac{2 s_{24}}{s_{34}^2}
+\frac{2 s_{14}}{s_{34}^2}
+\frac{2 s_{23}}{s_{34}^2}
+\frac{2 s_{12} s_{13}^2}{s_{134}^2 s_{34}^2}
+\frac{2 s_{12} s_{24}^2}{s_{234}^2 s_{34}^2}
-\frac{4 s_{12} s_{13}}{s_{134} s_{34}^2}
-\frac{4 s_{12} s_{24}}{s_{234} s_{34}^2}
\nonumber \\
& & \textstyle {} 
+\frac{2 s_{13}^2}{s_{134} s_{34}^2}
+\frac{2 s_{24}^2}{s_{234} s_{34}^2}
+\frac{2 s_{13}^2 s_{23}}{s_{134}^2 s_{34}^2}
+\frac{2 s_{14} s_{24}^2}{s_{234}^2 s_{34}^2}
-\frac{4 s_{13} s_{23}}{s_{134} s_{34}^2}
-\frac{4 s_{14} s_{24}}{s_{234} s_{34}^2}
+\frac{4 s_{12} s_{13} s_{24}}{s_{134} s_{234} s_{34}^2}
\nonumber \\
& & \textstyle {} 
+\frac{2 s_{13}^2 s_{24}}{s_{134}^2 s_{34}^2}
+\frac{2 s_{13} s_{24}^2}{s_{234}^2 s_{34}^2}
+\frac{6 s_{12}}{s_{13} s_{34}}
+\frac{6 s_{12}}{s_{24} s_{34}}
+\frac{4 s_{12} s_{13}}{s_{134}^2 s_{34}}
+\frac{4 s_{12} s_{24}}{s_{234}^2 s_{34}}
-\frac{6}{s_{34}}
\nonumber \\
& & \textstyle {} 
-\frac{12 s_{12}}{s_{134} s_{34}}
-\frac{12 s_{12}}{s_{234} s_{34}}
+\frac{6 s_{13}}{s_{134} s_{34}}
+\frac{6 s_{24}}{s_{234} s_{34}}
+\frac{2 s_{14}}{s_{13} s_{34}}
+\frac{2 s_{23}}{s_{24} s_{34}}
+\frac{4 s_{23}}{s_{13} s_{34}}
+\frac{4 s_{14}}{s_{24} s_{34}}
\nonumber \\
& & \textstyle {} 
+\frac{4 s_{13} s_{23}}{s_{134}^2 s_{34}}
+\frac{4 s_{14} s_{24}}{s_{234}^2 s_{34}}
-\frac{7 s_{23}}{s_{134} s_{34}}
-\frac{7 s_{14}}{s_{234} s_{34}}
+\frac{2 s_{12}^2}{s_{13} s_{234} s_{34}}
+\frac{2 s_{12}^2}{s_{134} s_{24} s_{34}}
\nonumber \\
& & \textstyle {} 
-\frac{s_{13}}{s_{234} s_{34}}
-\frac{s_{24}}{s_{134} s_{34}}
+\frac{4 s_{12} s_{13}}{s_{134} s_{234} s_{34}}
+\frac{4 s_{12} s_{24}}{s_{134} s_{234} s_{34}}
-\frac{2 s_{13}^2}{s_{134} s_{234} s_{34}}
-\frac{2 s_{24}^2}{s_{134} s_{234} s_{34}}
\nonumber \\
& & \textstyle {} 
+\frac{2 s_{12} s_{14}}{s_{13} s_{234} s_{34}}
+\frac{2 s_{12} s_{23}}{s_{134} s_{24} s_{34}}
+\frac{s_{14}^2}{s_{13} s_{234} s_{34}}
+\frac{s_{23}^2}{s_{134} s_{24} s_{34}}
+\frac{2 s_{12}^2}{s_{13} s_{24} s_{34}}
-\frac{8 s_{12}^2}{s_{134} s_{234} s_{34}}
\nonumber \\
& & \textstyle {} 
+\frac{s_{13}}{s_{24} s_{34}}
+\frac{s_{24}}{s_{13} s_{34}}
-\frac{2 s_{12} s_{13}}{s_{134} s_{24} s_{34}}
-\frac{2 s_{12} s_{24}}{s_{13} s_{234} s_{34}}
+\frac{s_{13}^2}{s_{134} s_{24} s_{34}}
+\frac{s_{24}^2}{s_{13} s_{234} s_{34}}
\nonumber \\
& & \textstyle {} 
+\frac{2 s_{12} s_{14}}{s_{13} s_{24} s_{34}}
+\frac{2 s_{12} s_{23}}{s_{13} s_{24} s_{34}}
+\frac{s_{14}^2}{s_{13} s_{24} s_{34}}
+\frac{s_{23}^2}{s_{13} s_{24} s_{34}}
+\frac{4 s_{13} s_{24}}{s_{134}^2 s_{34}}
+\frac{4 s_{13} s_{24}}{s_{234}^2 s_{34}}
\nonumber \\
& & \textstyle {} 
-\frac{2 s_{13} s_{23}}{s_{134} s_{24} s_{34}}
-\frac{2 s_{14} s_{24}}{s_{13} s_{234} s_{34}}
+\frac{s_{12} s_{34}}{s_{13} s_{134}^2}
+\frac{s_{12} s_{34}}{s_{234}^2 s_{24}}
+\frac{2 s_{34}}{s_{13} s_{134}}
+\frac{2 s_{34}}{s_{234} s_{24}}
-\frac{4 s_{13} s_{24}}{s_{134} s_{234} s_{34}}
\nonumber \\
& & \textstyle {} 
+\frac{s_{23} s_{34}}{s_{13} s_{134}^2}
+\frac{s_{14} s_{34}}{s_{234}^2 s_{24}}
+\frac{3 s_{34}}{s_{13} s_{234}}
+\frac{3 s_{34}}{s_{134} s_{24}}
+\frac{6 s_{12} s_{34}}{s_{13} s_{134} s_{234}}
+\frac{6 s_{12} s_{34}}{s_{134} s_{234} s_{24}}
-\frac{8 s_{34}}{s_{134} s_{234}}
\nonumber \\
& & \textstyle {} 
-\frac{3 s_{12} s_{34}}{s_{13} s_{134} s_{24}}
-\frac{3 s_{12} s_{34}}{s_{13} s_{234} s_{24}}
-\frac{s_{23} s_{34}}{s_{13} s_{134} s_{24}}
-\frac{s_{14} s_{34}}{s_{13} s_{234} s_{24}}
+\frac{s_{13} s_{34}}{s_{234}^2 s_{24}}
+\frac{s_{24} s_{34}}{s_{13} s_{134}^2}
\nonumber \\
& & \textstyle {} 
-\frac{3 s_{13} s_{34}}{s_{134} s_{234} s_{24}}
-\frac{3 s_{24} s_{34}}{s_{13} s_{134} s_{234}}
-\frac{3 s_{34}^2}{s_{13} s_{134} s_{234}}
-\frac{3 s_{34}^2}{s_{134} s_{234} s_{24}}
+\frac{s_{34}^2}{s_{13} s_{134} s_{24}}
+\frac{s_{34}^2}{s_{13} s_{234} s_{24}}
\nonumber \\
& & \textstyle {} 
+\frac{3 s_{12} s_{34}^2}{s_{13} s_{134} s_{234} s_{24}}
-\frac{s_{34}^3}{s_{13} s_{134} s_{234} s_{24}}
-\frac{4 s_{12}^2 s_{34}}{s_{13} s_{134} s_{234} s_{24}}
\nonumber \\
& & \textstyle {} + m_Q^2 \Big( 
  \frac{12}{s_{134}^2}
+\frac{12}{s_{234}^2}
-\frac{4}{s_{13} s_{134}}
-\frac{4}{s_{234} s_{24}}
-\frac{20}{s_{13} s_{234}}
-\frac{20}{s_{134} s_{24}}
+\frac{28}{s_{134} s_{234}}
\nonumber \\
& & \textstyle {} 
-\frac{4 (s_{12}+s_{14})^2}{s_{13} s_{234} s_{24}^2}
-\frac{4 (s_{12}+s_{23})^2}{s_{13}^2 s_{134} s_{24}}
-\frac{2 (2 s_{12}+s_{13}+2 s_{14})}{s_{234} s_{24}^2}
-\frac{2 (2 s_{12}+2 s_{23}+s_{24})}{s_{13}^2 s_{134}}
\nonumber \\
& & \textstyle {} 
+\frac{8 (s_{12}-s_{34})}{s_{13} s_{134} s_{24}}
+\frac{8 (s_{12}-s_{34})}{s_{13} s_{234} s_{24}}
+\frac{8 s_{13}^2}{s_{134}^2 s_{34}^2}
+\frac{8 s_{24}^2}{s_{234}^2 s_{34}^2}
-\frac{16 s_{13}}{s_{134} s_{34}^2}
-\frac{16 s_{24}}{s_{234} s_{34}^2}
\nonumber \\
& & \textstyle {} 
+\frac{8}{s_{13} s_{34}}
+\frac{8}{s_{24} s_{34}}
-\frac{20}{s_{134} s_{34}}
-\frac{20}{s_{234} s_{34}}
+\frac{8 s_{12}-12 s_{24}}{s_{13} s_{234} s_{34}}
+\frac{8 s_{12}-12 s_{13}}{s_{134} s_{24} s_{34}}
+\frac{8}{s_{34}^2} 
\nonumber \\
& & \textstyle {} 
-\frac{4 s_{12} s_{13}}{s_{134} s_{234} s_{24} s_{34}}
-\frac{4 s_{12} s_{24}}{s_{13} s_{134} s_{234} s_{34}}
-\frac{4 s_{23} (s_{12}+s_{23})}{s_{13} s_{134} s_{24} s_{34}}
-\frac{4 s_{14} (s_{12}+s_{14})}{s_{13} s_{234} s_{24} s_{34}}
+\frac{6}{s_{13} s_{24}} 
\nonumber \\
& & \textstyle {} 
-\frac{4 (s_{13}+s_{14})}{s_{234} s_{24} s_{34}}
-\frac{4 (s_{23}+s_{24})}{s_{13} s_{134} s_{34}}
+\frac{4 (s_{12}+4 s_{13}+s_{23}+s_{24})}{s_{134}^2 s_{34}}
+\frac{4 (s_{12}+s_{13}+s_{14}+4 s_{24})}{s_{234}^2 s_{34}}
\nonumber \\
& & \textstyle {} 
+\frac{4 (s_{12}+s_{13}+s_{14}+s_{34})}{s_{234}^2 s_{24}}
+\frac{4 (s_{12}+s_{23}+s_{24}+s_{34})}{s_{13} s_{134}^2}
+\frac{8 s_{12}-4 (s_{14}+s_{23})}{s_{13} s_{24} s_{34}}
+\frac{16 s_{13} s_{24}}{s_{134} s_{234} s_{34}^2}
\nonumber \\
& & \textstyle {} 
+\frac{16 (s_{13}+s_{24}-2 s_{12})}{s_{134} s_{234} s_{34}}
-\frac{2 (2 s_{12}+2 s_{14}+s_{23}+s_{34})}{s_{13} s_{24}^2}
-\frac{2 (2 s_{12}+s_{14}+2 s_{23}+s_{34})}{s_{13}^2 s_{24}}
\nonumber \\
& & \textstyle {} 
+\frac{-20 s_{12}+8 s_{13}+14 s_{34}}{s_{134} s_{234} s_{24}}
+\frac{-20 s_{12}+8 s_{24}+14 s_{34}}{s_{13} s_{134} s_{234}}
+\frac{8 s_{12}^2-14 s_{12} s_{34}+8 s_{34}^2}{s_{13} s_{134} s_{234} s_{24}}
\Big)
\nonumber \\
& & \textstyle {} + m_Q^4 \Big( 
  \frac{16}{s_{13} s_{134}^2}
+ \frac{16}{s_{234}^2 s_{24}}
+ \frac{4}{s_{13}^2 s_{134}}
+ \frac{4}{s_{234} s_{24}^2}
+ \frac{4}{s_{13} s_{24}^2}
+ \frac{4}{s_{13}^2 s_{24}}
\nonumber \\
& & \textstyle {} 
+ \frac{16}{s_{134}^2 s_{34}}
+ \frac{16}{s_{234}^2 s_{34}}
- \frac{16 (s_{12}+s_{14})}{s_{13} s_{234} s_{24}^2}
- \frac{16  (s_{12}+s_{23})}{s_{13}^2 s_{134} s_{24}}
- \frac{16}{s_{13} s_{134} s_{234}}
- \frac{16}{s_{134} s_{234} s_{24}}
\nonumber \\
& & \textstyle {} 
- \frac{16 s_{23}}{s_{13} s_{134} s_{24} s_{34}}
- \frac{16 s_{13}}{s_{134} s_{234} s_{24} s_{34}}
- \frac{16 s_{14}}{s_{13} s_{234} s_{24} s_{34}}
- \frac{16 s_{24}}{s_{13} s_{134} s_{234} s_{34}}
\nonumber \\
& & \textstyle {} 
+ \frac{4 (s_{12}+s_{13}+s_{14})}{s_{234}^2 s_{24}^2}
+ \frac{4 (s_{12}+s_{23}+s_{24})}{s_{13}^2 s_{134}^2}
- \frac{12 s_{34}}{s_{13} s_{134} s_{234} s_{24}}
+ \frac{4 (s_{12}+s_{14}+s_{23}+s_{34})}{s_{13}^2 
s_{24}^2} \Big)
\nonumber \\
& & \textstyle {} + m_Q^6 \left( 
  \frac{16}{ s_{13}^2 s_{134}^2 } 
+ \frac{16}{ s_{13}^2 s_{24}^2 } 
+ \frac{16}{ s_{24}^2 s_{234}^2 }
\right) \Big) + \cO( \epsilon ) \,,
\label{eq::A40}
\\[3ex]
\tilde{A}^0_4 \! \left ( 1_Q, 3_g, 4_g, 2_{\bar{Q}} \right)
& = & \textstyle \frac{ 1 }{ s_{1342} + 4 m_Q^2 } \Big(
-\frac{2}{s_{13}}
-\frac{2}{s_{24}}
-\frac{1}{s_{23}}
-\frac{1}{s_{14}}
+\frac{2}{s_{134}}
+\frac{2}{s_{234}}
+\frac{3 s_{12}}{2 s_{13} s_{14}}
+\frac{3 s_{12}}{2 s_{23} s_{24}}
\nonumber \\
& & \textstyle {}
+\frac{s_{12}}{2 s_{13} s_{134}}
+\frac{s_{12}}{2 s_{234} s_{24}}
+\frac{s_{12}}{2 s_{23} s_{234}}
+\frac{s_{12}}{2 s_{134} s_{14}}
+\frac{2 (s_{12}+s_{13}+s_{14})}{s_{234}^2}
+\frac{2 (s_{12}+s_{23}+s_{24})}{s_{134}^2}
\nonumber \\
& & \textstyle {}
+\frac{2 s_{12}^2}{s_{13} s_{14} s_{23}}
+\frac{2 s_{12}^2}{s_{14} s_{23} s_{24}}
+\frac{2 s_{12}^2}{s_{13} s_{14} s_{24}}
+\frac{2 s_{12}^2}{s_{13} s_{23} s_{24}}
+\frac{2 s_{12}^3}{s_{13} s_{14} s_{23} s_{24}}
+\frac{4 (s_{12}-s_{34})}{s_{134} s_{234}}
\nonumber \\
& & \textstyle {}
+\frac{2 s_{12}^3}{s_{13} s_{134} s_{23} s_{234}}
+\frac{2 s_{12}^3}{s_{134} s_{14} s_{234} s_{24}}
-\frac{2 s_{12}+2 s_{13}+s_{24}-s_{34}}{s_{14} s_{234}}
-\frac{2 s_{12}+s_{13}+2 s_{24}-s_{34}}{s_{134} s_{23}}
\nonumber \\
& & \textstyle {}
+\frac{7 s_{12} s_{13}}{2 s_{14} s_{23} s_{234}}
+\frac{7 s_{12} s_{24}}{2 s_{134} s_{14} s_{23}}
+\frac{7 s_{12} s_{23}}{2 s_{13} s_{134} s_{24}}
+\frac{7 s_{12} s_{14}}{2 s_{13} s_{234} s_{24}}
\nonumber \\
& & \textstyle {}
+\frac{3 s_{12} s_{14}}{2 s_{13} s_{23} s_{234}}
+\frac{3 s_{12} s_{23}}{2 s_{134} s_{14} s_{24}}
+\frac{3 s_{12} s_{13}}{2 s_{14} s_{234} s_{24}}
+\frac{3 s_{12} s_{24}}{2 s_{13} s_{134} s_{23}}
\nonumber \\
& & \textstyle {}
+\frac{s_{12} s_{13}}{2 s_{14} s_{23} s_{24}}
+\frac{s_{12} s_{14}}{2 s_{13} s_{23} s_{24}}
+\frac{s_{12} s_{23}}{2 s_{13} s_{14} s_{24}}
+\frac{s_{12} s_{24}}{2 s_{13} s_{14} s_{23}}
\nonumber \\
& & \textstyle {}
+\frac{3 (2 s_{12}+s_{14}+s_{23})}{s_{13} s_{24}}
+\frac{3 (2 s_{12}+s_{13}+s_{24})}{s_{14} s_{23}}
-\frac{s_{13}+s_{14}-s_{34}}{s_{23} s_{234}}
-\frac{s_{23}+s_{24}-s_{34}}{s_{134} s_{14}}
\nonumber \\
& & \textstyle {}
-\frac{s_{13}+s_{14}-2 s_{34}}{s_{234} s_{24}}
-\frac{s_{23}+s_{24}-2 s_{34}}{s_{13} s_{134}}
+\frac{3 s_{12} s_{34}}{2 s_{13} s_{134}^2}
+\frac{3 s_{12} s_{34}}{2 s_{134}^2 s_{14}}
+\frac{3 s_{12} s_{34}}{2 s_{23} s_{234}^2}
+\frac{3 s_{12} s_{34}}{2 s_{234}^2 s_{24}}
\nonumber \\
& & \textstyle {}
+\frac{(s_{13}+s_{14}) s_{34}}{s_{23} s_{234}^2}
+\frac{(s_{23}+s_{24}) s_{34}}{s_{134}^2 s_{14}}
+\frac{(s_{13}+s_{14}) s_{34}}{s_{234}^2 s_{24}}
+\frac{(s_{23}+s_{24}) s_{34}}{s_{13} s_{134}^2}
\nonumber \\
& & \textstyle {}
+\frac{s_{12} s_{23} s_{34}}{2 s_{13} s_{134} s_{14} s_{24}}
+\frac{s_{12} s_{14} s_{34}}{2 s_{13} s_{23} s_{234} s_{24}}
+\frac{s_{12} s_{13} s_{34}}{2 s_{14} s_{23} s_{234} s_{24}}
+\frac{s_{12} s_{24} s_{34}}{2 s_{13} s_{134} s_{14} s_{23}}
\nonumber \\
& & \textstyle {}
+\frac{(3 s_{12}-s_{13}-2 s_{34}) s_{34}}{s_{134} s_{234} s_{24}}
+\frac{(3 s_{12}-s_{24}-2 s_{34}) s_{34}}{s_{13} s_{134} s_{234}}
+\frac{(3 s_{12}+s_{13}-s_{34}) s_{34}}{s_{134} s_{23} s_{234}}
+\frac{(3 s_{12}+s_{24}-s_{34}) s_{34}}{s_{134} s_{14} s_{234}}
\nonumber \\
& & \textstyle {}
+\frac{s_{12} s_{34}^2}{2 s_{13} s_{134}^2 s_{14}}
+\frac{s_{12} s_{34}^2}{2 s_{23} s_{234}^2 s_{24}}
+\frac{-2 s_{12}+s_{13}-2 s_{23}+2 s_{34}}{s_{134} s_{24}}
+\frac{-2 s_{12}-2 s_{14}+s_{24}+2 s_{34}}{s_{13} s_{234}}
\nonumber \\
& & \textstyle {}
+\frac{s_{12} (2 s_{12}+s_{34})}{s_{13} s_{134} s_{23}}
+\frac{s_{12} (2 s_{12}+s_{34})}{s_{14} s_{234} s_{24}}
+\frac{s_{12} (2 s_{12}+s_{34})}{s_{13} s_{23} s_{234}}
+\frac{s_{12} (2 s_{12}+s_{34})}{s_{134} s_{14} s_{24}}
\nonumber \\
& & \textstyle {}
+\frac{4 s_{12}^2+s_{13}^2-3 s_{12} s_{34}-s_{13} s_{34}+s_{34}^2}{s_{14} 
s_{23} s_{234}}
+\frac{4 s_{12}^2+s_{24}^2-3 s_{12} s_{34}-s_{24} s_{34}+s_{34}^2}{s_{134} 
s_{14} s_{23}}
\nonumber \\
& & \textstyle {}
+\frac{4 s_{12}^2+s_{14}^2-3 s_{12} s_{34}-s_{14} s_{34}+s_{34}^2}{s_{13} 
s_{234} s_{24}}
+\frac{4 s_{12}^2+s_{23}^2-3 s_{12} s_{34}-s_{23} s_{34}+s_{34}^2}{s_{13} 
s_{134} s_{24}}
\nonumber \\
& & \textstyle {}
+\frac{2 s_{12}^3-4 s_{12}^2 s_{34}+3 s_{12} s_{34}^2-s_{34}^3}{s_{134} s_{14} 
s_{23} s_{234}}
+\frac{2 s_{12}^3-4 s_{12}^2 s_{34}+3 s_{12} s_{34}^2-s_{34}^3}{s_{13} s_{134} 
s_{234} s_{24}}
\nonumber \\
& & \textstyle {} + m_Q^2 \Big( 
\frac{7}{s_{13} s_{134}}
+\frac{7}{s_{234} s_{24}}
+\frac{7}{s_{23} s_{234}}
+\frac{7}{s_{134} s_{14}}
-\frac{7}{s_{13} s_{14}}
-\frac{7}{s_{23} s_{24}}
\nonumber \\
& & \textstyle {}
-\frac{12}{s_{134} s_{23}}
-\frac{12}{s_{14} s_{234}}
-\frac{12}{s_{13} s_{234}}
-\frac{12}{s_{134} s_{24}}
+\frac{8}{s_{134}^2}
+\frac{8}{s_{234}^2}
+\frac{16}{s_{134} s_{234}}
\nonumber \\
& & \textstyle {}
-\frac{3 s_{12}}{s_{13} s_{134} s_{23}}
-\frac{3 s_{12}}{s_{14} s_{234} s_{24}}
-\frac{3 s_{12}}{s_{13} s_{23} s_{234}}
-\frac{3 s_{12}}{s_{134} s_{14} s_{24}}
\nonumber \\
& & \textstyle {}
+\frac{8}{s_{14} s_{23}}
+\frac{8}{s_{13} s_{24}}
-\frac{s_{12}}{s_{13} s_{14} s_{23}}
-\frac{s_{12}}{s_{14} s_{23} s_{24}}
-\frac{s_{12}}{s_{13} s_{14} s_{24}}
-\frac{s_{12}}{s_{13} s_{23} s_{24}}
\nonumber \\
& & \textstyle {}
-\frac{4 (s_{12}+s_{13})^2}{s_{14} s_{23}^2 s_{234}}
-\frac{4 (s_{12}+s_{24})^2}{s_{134} s_{14}^2 s_{23}}
-\frac{4 (s_{12}+s_{14})^2}{s_{13} s_{234} s_{24}^2}
-\frac{4 (s_{12}+s_{23})^2}{s_{13}^2 s_{134} s_{24}}
\nonumber \\
& & \textstyle {}
-\frac{2 (2 s_{12}+2 s_{13}+s_{14})}{s_{23}^2 s_{234}}
-\frac{2 (2 s_{12}+s_{13}+2 s_{14})}{s_{234} s_{24}^2}
-\frac{2 (2 s_{12}+2 s_{23}+s_{24})}{s_{13}^2 s_{134}}
-\frac{2 (2 s_{12}+s_{23}+2 s_{24})}{s_{134} s_{14}^2}
\nonumber \\
& & \textstyle {}
+\frac{5 s_{12}-8 s_{34}}{s_{134} s_{14} s_{23}}
+\frac{5 s_{12}-8 s_{34}}{s_{14} s_{23} s_{234}}
+\frac{5 s_{12}-8 s_{34}}{s_{13} s_{134} s_{24}}
+\frac{5 s_{12}-8 s_{34}}{s_{13} s_{234} s_{24}}
+\frac{s_{34}^2}{s_{13} s_{134}^2 s_{14}}
+\frac{s_{34}^2}{s_{23} s_{234}^2 s_{24}}
\nonumber \\
& & \textstyle {}
-\frac{4 (s_{12}-2 s_{34})}{s_{13} s_{134} s_{234}}
-\frac{4 (s_{12}-2 s_{34})}{s_{134} s_{234} s_{24}}
-\frac{4 (s_{12}-2 s_{34})}{s_{134} s_{14} s_{234}}
-\frac{4 (s_{12}-2 s_{34})}{s_{134} s_{23} s_{234}}
\nonumber \\
& & \textstyle {}
-\frac{2 (4 s_{12}+s_{13}+s_{14}+s_{34})}{s_{23} s_{234} s_{24}}
-\frac{2 (4 s_{12}+s_{23}+s_{24}+s_{34})}{s_{13} s_{134} s_{14}}
-\frac{2 (2 s_{12}+2 s_{14}+s_{23}+s_{34})}{s_{13} s_{24}^2}
\nonumber \\
& & \textstyle {}
-\frac{2 (2 s_{12}+2 s_{13}+s_{24}+s_{34})}{s_{14} s_{23}^2}
-\frac{2 (2 s_{12}+s_{14}+2 s_{23}+s_{34})}{s_{13}^2 s_{24}}
-\frac{2 (2 s_{12}+s_{13}+2 s_{24}+s_{34})}{s_{14}^2 s_{23}}
\nonumber \\
& & \textstyle {}
-\frac{s_{12} (4 s_{12}+4 s_{13}+3 s_{34})}{s_{14} s_{23} s_{234} 
s_{24}}
-\frac{s_{12} (4 s_{12}+4 s_{24}+3 s_{34})}{s_{13} s_{134} s_{14} 
s_{23}}
-\frac{s_{12} (4 s_{12}+4 s_{14}+3 s_{34})}{s_{13} s_{23} s_{234} 
s_{24}}
\nonumber \\
& & \textstyle {}
-\frac{s_{12} (4 s_{12}+4 s_{23}+3 s_{34})}{s_{13} s_{134} s_{14} 
s_{24}}
+\frac{4 s_{12}+4 s_{13}+4 s_{14}+5 s_{34}}{s_{23} s_{234}^2}
+\frac{4 s_{12}+4 s_{13}+4 s_{14}+5 s_{34}}{s_{234}^2 s_{24}}
\nonumber \\
& & \textstyle {}
+\frac{4 s_{12}+4 s_{23}+4 s_{24}+5 s_{34}}{s_{13} s_{134}^2}
+\frac{4 s_{12}+4 s_{23}+4 s_{24}+5 s_{34}}{s_{134}^2 s_{14}}
+\frac{2 \left(4 s_{12}^2-s_{12} s_{34}+s_{34}^2\right)}{s_{13} s_{14} s_{23} 
s_{24}}
\nonumber \\
& & \textstyle {}
+\frac{2 \left(4 s_{12}^2+s_{12} s_{34}+s_{34}^2\right)}{s_{13} s_{134} 
s_{23} s_{234}}
+\frac{2 \left(4 s_{12}^2+s_{12} s_{34}+s_{34}^2\right)}{s_{134} s_{14} s_{234} 
s_{24}}
+\frac{8 s_{12}^2-14 s_{12} s_{34}+8 s_{34}^2}{s_{134} s_{14} s_{23} 
s_{234}}
\nonumber \\
& & \textstyle {}
+\frac{8 s_{12}^2-14 s_{12} s_{34}+8 s_{34}^2}{s_{13} s_{134} 
s_{234} s_{24}}
\Big)
\nonumber \\
& & \textstyle {} + m_Q^4 \Big( 
 \frac{24}{s_{13} s_{134}^2}
+\frac{24}{s_{234}^2 s_{24}}
+\frac{24}{s_{134}^2 s_{14}}
+\frac{24}{s_{23} s_{234}^2}
+\frac{4}{s_{13}^2 s_{134}}
+\frac{4}{s_{234} s_{24}^2}
\nonumber \\
& & \textstyle {} 
+\frac{4}{s_{134} s_{14}^2}
+\frac{4}{s_{23}^2 s_{234}}
+\frac{4}{s_{14} s_{23}^2}
+\frac{4}{s_{14}^2 s_{23}}
+\frac{4}{s_{13} s_{24}^2}
+\frac{4}{s_{13}^2 s_{24}}
\nonumber \\
& & \textstyle {} 
-\frac{16}{s_{13} s_{134} s_{234}}
-\frac{16}{s_{134} s_{14} s_{234}}
-\frac{16}{s_{134} s_{23} s_{234}}
-\frac{16}{s_{134} s_{234} s_{24}}
+\frac{12 s_{34}}{s_{13} s_{14} s_{23} s_{24}}
\nonumber \\
& & \textstyle {} 
-\frac{16 s_{12}}{s_{13} s_{134} s_{14} s_{23}}
-\frac{16 s_{12}}{s_{14} s_{23} s_{234} s_{24}}
-\frac{16 s_{12}}{s_{13} s_{134} s_{14} s_{24}}
-\frac{16 s_{12}}{s_{13} s_{23} s_{234} s_{24}}
\nonumber \\
& & \textstyle {} 
+\frac{4 (s_{12}+s_{13}+s_{14})}{s_{23}^2 s_{234}^2}
+\frac{4 (s_{12}+s_{23}+s_{24})}{s_{134}^2 s_{14}^2}
+\frac{4 (s_{12}+s_{13}+s_{14})}{s_{234}^2 s_{24}^2}
+\frac{4 (s_{12}+s_{23}+s_{24})}{s_{13}^2 s_{134}^2}
\nonumber \\
& & \textstyle {} 
-\frac{16 (s_{12}+s_{13})}{s_{14} s_{23}^2 s_{234}}
-\frac{16 (s_{12}+s_{24})}{s_{134} s_{14}^2 s_{23}}
-\frac{16 (s_{12}+s_{14})}{s_{13} s_{234} s_{24}^2}
-\frac{16 (s_{12}+s_{23})}{s_{13}^2 s_{134} s_{24}}
\nonumber \\
& & \textstyle {} 
-\frac{12 s_{34}}{s_{13} s_{134} s_{23} s_{234}}
-\frac{12 s_{34}}{s_{134} s_{14} s_{234} s_{24}}
-\frac{12 s_{34}}{s_{134} s_{14} s_{23} s_{234}}
-\frac{12 s_{34}}{s_{13} s_{134} s_{234} s_{24}}
\nonumber \\
& & \textstyle {} 
+\frac{4 (s_{12}+s_{14}+s_{23}+s_{34})}{s_{13}^2 s_{24}^2}
+\frac{4 (s_{12}+s_{13}+s_{24}+s_{34})}{s_{14}^2 s_{23}^2}
\nonumber \\
& & \textstyle {} 
+\frac{8 (s_{12}+s_{13}+s_{14}+s_{34})}{s_{23} s_{234}^2 s_{24}}
+\frac{8 (s_{12}+s_{23}+s_{24}+s_{34})}{s_{13} s_{134}^2 s_{14}}
\Big)
\nonumber \\
& & \textstyle {} + m_Q^6 \Big(
\frac{16}{s_{13}^2 s_{134}^2}
+\frac{16}{s_{234}^2 s_{24}^2}
+\frac{16}{s_{134}^2 s_{14}^2}
+\frac{16}{s_{23}^2 s_{234}^2}
+\frac{16}{s_{14}^2 s_{23}^2}
+\frac{16}{s_{13}^2 s_{24}^2}
\nonumber \\
& & \textstyle {} 
+\frac{32}{s_{13} s_{134}^2 s_{14}}
+\frac{32}{s_{23} s_{234}^2 s_{24}}
\Big) \Big)
+ \cO( \epsilon ) \,.
\label{eq::A40t}
\end{eqnarray}
\end{document}